\documentclass[prb,aps,twocolumn,superscriptaddress,floatfix,citeautoscript,eqsecnum,longbibliography]{revtex4-2}

\usepackage{graphicx,rotating,subfigure,amsmath,amsfonts,amssymb,delarray,color}

\usepackage{dsfont}
\usepackage[T1]{fontenc}
\usepackage{multirow}
\usepackage{comment}
\usepackage{hyperref}

\newcommand{\e}{\text{e}}

\def\l{\left}
\def\r{\right}
\def\nn{\nonumber}
\newcommand{\eq}[1]{(\ref{eq:#1})}
\newcommand{\dr}[1]{\widetilde{#1}}

\newcommand{\aux}{\mathfrak{a}}
\newcommand{\gC}{\mathcal{C}}

\newcommand{\shL}{\mathcal{L}}
\newcommand{\ii}{\mathrm{i}}

\newcommand{\ka}{\widetilde{K}}

\newcommand{\lna}{\ln A}
\newcommand{\lnb}{\ln \bar{A}}

\begin{document}

\title{Analytical results for the low-temperature Drude weight of the XXZ spin chain}
\author{Andrew Urichuk}
\email{urichuka@myumanitoba.ca}
\affiliation{Department of Physics and Astronomy, University of Manitoba, Winnipeg R3T 2N2, Canada}
\affiliation{Fakult\"at f\"ur Mathematik und Naturwissenschaften, Bergische Universit\"at Wuppertal, 42097 Wuppertal, Germany}
\author{Jesko Sirker}
\affiliation{Department of Physics and Astronomy, University of Manitoba, Winnipeg R3T 2N2, Canada}
\affiliation{Manitoba Quantum Institute, University of Manitoba, Winnipeg R3T 2N2, Canada}
\author{Andreas Kl\"umper}
\affiliation{Fakult\"at f\"ur Mathematik und Naturwissenschaften, Bergische Universit\"at Wuppertal, 42097 Wuppertal, Germany}

\begin{abstract}
The spin-$1/2$ XXZ chain is an integrable lattice model and parts of
its spin current can be protected by local conservation laws for
anisotropies $-1<\Delta<1$. In this case, the Drude weight $D(T)$ is
non-zero at finite temperatures $T$. Here we obtain analytical results
for $D(T)$ at low temperatures for zero external magnetic field and
anisotropies $\Delta=\cos(n\pi/m)$ with $n,m$ coprime integers, using
the thermodynamic Bethe ansatz. We show that to leading orders
$D(T)=D(0)-a(\Delta)T^{2K-2}-b_1(\Delta)T^2$ where $K$ is the
Luttinger parameter and the prefactor $a(\Delta)$, obtained in closed
form, has a fractal structure as function of anisotropy $\Delta$. The
prefactor $b_1(\Delta)$, on the other hand, does not have a fractal
structure and can be obtained in a standard field-theoretical
approach. Including both temperature corrections, we obtain an
analytic result for the low-temperature asymptotics of the Drude
weight in the entire regime $-1<\Delta=\cos(n\pi/m)<1$.
\end{abstract}
\maketitle

\section{Introduction}
One-dimensional integrable quantum systems can exhibit unusual
transport properties~\cite{shastry_twisted_1990, zotos_transport_1997,
castella_integrability_1995,zotos_finite_1999,sakai_non-dissipative_2003,klumper_thermal_2002,sirker_diffusion_2009,sirker_conservation_2011,prosen_exact_2011,kinoshita_quantum_2006},
which are, in principle, exactly computable by the Bethe ansatz. Bethe
ansatz and related techniques have been used to successfully
investigate many different aspects of these
models~\cite{zvyagin_momentum_1991,zotos_finite_1999,sakai_non-dissipative_2003,klumper_thermal_2002,boos_factorization_2006,benz_finite_2005,aufgebauer_complete_2010,prosen_exact_2011,GohmannLecture,GohmannKozlowski,BabenkoGohmann}. More
recently, the advent of generalized hydrodynamics
(GHD)~\cite{castro-alvaredo_emergent_2016,bertini_transport_2016,doyon_drude_2017,ilievski_microscopic_2017}
has provided a new avenue to explore their transport properties. GHD
combines the thermodynamic Bethe ansatz
(TBA)~\cite{takahashi_thermodynamics_1999} with a continuity equation
and an assumed form for the current operator, which has been
demonstrated to be exact in
Refs.~\cite{borsi_current_2020,pozsgay_current_2020,cubero_generalized_2020}
on the level of form factors. By use of a projective
formalism~\cite{doyon_diffusion_2019}, formulas for the Drude
weight~\cite{doyon_drude_2017,ilievski_ballistic_2017} and diffusion
coefficients
\cite{de_nardis_hydrodynamic_2018,de_nardis_anomalous_2019,ilievski_superdiffusion_2018}
have been determined from GHD. For the XXZ spin chain, these formulas
for the Drude weight do agree with the result derived 20 years earlier
by Zotos \cite{zotos_finite_1999} which relies only on the TBA. So
far, however, analytical results for the Drude weight are only
available at zero temperature \cite{shastry_twisted_1990} and at
infinite temperature
\cite{prosen_exact_2011,prosen_families_2013,pereira_exactly_2014}. For finite temperatures, on the other hand, 
these approaches lead to a set of equations that have only been solved
numerically up to now. The objective of this paper is to obtain
closed-form expressions for the Drude weight of the XXZ chain at low
temperatures and zero magnetic field.

The Hamiltonian of the XXZ spin chain is given by
\begin{equation}
\label{Ham}
H=\sum_{l=1}^M \left\{\frac{J}{4}\left(\sigma^x_l\sigma^x_{l+1} +\sigma^y_l\sigma^y_{l+1} +\Delta\sigma^z_l\sigma^z_{l+1}\right)-\frac{h}{2}\sigma^z_l\right\} \, ,
\end{equation}
where $J$ is the exchange constant, $M$ the number of lattice sites,
$\sigma^{x,y,z}$ are Pauli matrices, and $\Delta=\cos(\gamma)$ the
anisotropy parametrized by $\gamma$. The magnetic field is denoted by $h$ and we use periodic boundary conditions. The XXZ chain is a quantum integrable model which has a
family of commuting transfer matrices, $[T(\theta),T(\theta')]=0$ with
$\theta$, $\theta'$ being spectral parameters. These transfer matrices
generate an infinite set of local conserved charges $Q_n$, which are
obtained by
\begin{equation}%consistent with previous paper
\label{Charge}
Q_n = \frac{d^n}{d\theta^n} \ln T(\theta)\bigg|_{\theta =0}.
\end{equation}
In particular, $Q_1\propto H$ and $Q_2\propto J_E$ where $J_E$ is the
energy current operator. Because $J_E$ itself is conserved, the
thermal conductivity in linear response can be calculated
straightforwardly from a generalized Gibbs ensemble (GGE), which
includes the higher conserved charges
\cite{sakai_non-dissipative_2003,zotos_tba_2017}. 
Spin transport in the XXZ chain, on the other hand, is much more
complicated to investigate because the spin current operator $J_0$ is
not conserved
\cite{zotos_transport_1997}.

%The main objective of this paper is to obtain a low-temperature
%asymptotic formula for the Drude weight of the XXZ chain \eqref{Ham}
%at zero magnetic field, $h=0$. 
Our paper is organized as follows: in Sec.~\ref{Drude} we present
different approaches to calculate spin transport in linear
response. In Sec.~\ref{NLIE} we review the asymptotic analysis of a
system of non-linear integral equations for the free energy of the XXZ
chain. We then show in Sec.~\ref{Fermi} that a similar asymptotic
analysis can also be performed for the Fermi weight. Using the derived
asymptotic formulas for the Fermi weight, we then obtain in
Sec.~\ref{Asymp1} an analytic low-temperature expansion of $D(T)$ at
the simple roots of unity, i.e., anisotropies $\gamma=\pi/m$. The
generalization to anisotropies $\gamma=n\pi/m$ is discussed in
Sec.~\ref{Asymp2}.
%The low-temperature results show a
%fractal character as a function of anisotropy $\gamma$ which is
%analyzed further in Sec.~\ref{Fractal}. 
By including the leading temperature correction with integer scaling
dimension known from field theory and comparing to a numerical
solution of the TBA equations, we show in Sec.~\ref{Numerics} that the
obtained analytical asymptotics describes the Drude weight correctly
at low temperatures in the entire regime
$-1<\Delta=\cos(n\pi/m)<1$. The final section is devoted to a short
summary and some conclusions.

\section{Drude weight}
\label{Drude}
The spin Drude weight can be defined in the following two equivalent
ways. Firstly, using the Kubo formalism the spin conductivity
$\sigma(\omega)$ as a function of frequency $\omega$ can be expressed in
linear response as
\begin{equation}
\label{eq:kuboFm}
\sigma(\omega) = \frac{\ii}{\omega} \left[ \frac{ \langle E_\text{kin}\rangle}{N} + \langle J_0, J_0 \rangle_\text{ret}(\omega) \right].
\end{equation}
Here $\langle E_{\text{kin}}\rangle$ is the expectation value of the
kinetic energy of the Hamiltonian and $\langle J_0
,J_0\rangle_{\text{ret}}(\omega)$ denotes the retarded current-current
correlation function. The current operator is given by $J_0 = \sum_l
j_l$ with
\begin{equation}
\label{Current}
j_l = \ii J\left(\sigma^+_l\sigma^-_{l+1} - h.c.\right) 
\end{equation}
and $\sigma^\pm =(\sigma^x\pm \ii \sigma^y)/2$. The real part of the
conductivity can then be written as
\begin{equation}
\sigma'(\omega)=2 \pi D(T) \delta(\omega) + \sigma_\text{reg}(T,\omega),
\end{equation}
where a non-zero Drude weight $D(T)$ implies an infinite dc
conductivity at a given temperature $T$ and
$\sigma_\text{reg}(T,\omega)$ is the regular part of the conductivity. 

A second expression relates the Drude weight directly to the
real-time current-current correlator and the conserved charges $Q_k$
of the XXZ chain as
\begin{equation}
\label{Mazur}
2T D =\lim_{t \to \infty} \lim_{N \to \infty} \frac{\langle J_0(0) J_0(t) \rangle}{N} %,\nn\\
 \geq \lim_{N \to \infty} \sum_k \frac{|\langle J_0 Q_k \rangle |^2}{N\langle Q^\dag_k Q_k\rangle}.
\end{equation}
The first equation gives a nice physical interpretation of the Drude
weight. It shows that the Drude weight measures the part of the
current with no decay in time. In the second inequality, a set
of conserved charges $[H,Q_k]=0$ with $\langle Q_k Q_l\rangle
=\delta_{kl}$ is inserted, making the projection onto the conserved
part explicit. If only some of the conserved charges are taken into
account then a lower bound, the so-called Mazur bound, is obtained
\cite{mazur_non-ergodicity_1969}. The spin current \eqref{Current} is
odd under spin-flip symmetry $\sigma^z \to -\sigma^z$ while all the
conserved charges obtained from Eq.~\eqref{Charge} are even when
$h=0$. These local charges therefore have no overlap with the current
operator at zero magnetic field. In addition, there exist, however,
also quasi-local charges for $\gamma=n\pi/m\in \mathbb{Q}$ with $n,m$
coprime obtained from transfer matrices when a general highest weight
representation in auxiliary space is chosen instead of the standard
spin-$1/2$ representation
\cite{prosen_exact_2011,prosen_families_2013,pereira_exactly_2014}. Completeness of 
the set of local and quasi-local conserved charges is believed to be a
consequence of the underlying quantum
group~\cite{ilievski_equilibrium_2019}, and taking all the symmetries
into account turns the Mazur bound in Eq.~\eqref{Mazur} into an
equality~\cite{mazur_non-ergodicity_1969}.

For anisotropies $\gamma/\pi = n/m \in \mathbb{Q}$ with $n,m$ coprime
the Drude weight can be computed by multiple
methods. The first amounts to a generalization of the Kohn
formula~\cite{kohn_theory_1964} to finite
temperatures~\cite{castella_integrability_1995}. By threading a static
magnetic flux $\Phi$ through an XXZ ring and relating the now shifted eigenenergies
$\epsilon_n$ to the Kubo formula~\eq{kuboFm} one finds
\begin{equation}
\label{Kohn}
D = \frac{1}{2NZ} \sum_n \e^{-\epsilon_n/T} \frac{\partial^2 \epsilon_n(\Phi)}{\partial \Phi^2}\Big\rvert_{\Phi=0},
\end{equation}
with $Z$ being the partition function. Using the TBA
formalism~\cite{takahashi_thermodynamics_1999}, this method has been
used in Ref.~\cite{zotos_finite_1999} to obtain the explicit formula
\eqref{eq:zotosForm} below, which we will call Zotos' formula from here
on. The second method is based on GHD and the Mazur bound. Assuming
that the set of charges is complete, Zotos' formula is reproduced.
%Through a projective method, GHD makes use
%of the Mazur bound to determine the Drude
%weight~\cite{doyon_drude_2017,ilievski_ballistic_2017}, which is again
%equivalent to the assumption of a complete set of
%charges~\cite{urichuk_spin_2018}. The projection relies on TBA to
%define a dressing transformation
%\begin{equation}
%\dr{q}_n = q_n - \sum_m \int d\mu \, \mathcal{K}_{n m}(\theta-\mu) \vartheta_m(\mu) \sigma_m  \dr{q}_m(\mu),
%\end{equation}
%where $\mathcal{K}$ is the derivative of the scattering kernel of the model, $\vartheta$ is the Fermi weight, $q_n$ is the bare charge eigenvalue, and $\dr{q}_n$ is the GHD-dressed value. This method relies on the form of the conserved current, which was recently proven by studying the form factors~\cite{borsi_current_2019,pozsgay_current_2020,cubero_generalized_2020}. 
The third method is to explicitly construct the conserved charges. The
difficulty then is to calculate overlaps of these charges with the
current operator as well as the norm of these charges at finite
temperatures. So far, this program has only been carried out explicitly at
infinite temperatures
\cite{prosen_exact_2011,prosen_families_2013,pereira_exactly_2014}.
%For
%finite temperatures this is quite difficult and still
%ongoing\footnote{Unpublished notes from Christina Ballnus}, however
%infinite temperature calculations have been carried out using the
%quasi-local
%charges~\cite{prosen_exact_2011,prosen_families_2013,pereira_exactly_2014}.

The result of the first two methods is Zotos'
formula~\cite{zotos_finite_1999,benz_finite_2005} which can be
expressed in terms of hole/particle ratios $\eta_\ell =
\rho^h_\ell/\rho_\ell$ and factors $\sigma_\ell=
\frac{\dr{\epsilon_\ell}}{|\dr{\epsilon_\ell}|}=\pm 1$, where
$\dr{\epsilon_\ell}$ are the dressed quasi-particle energies. Zotos'
formula for the spin Drude weight can be written in terms of only the
last two strings (particle and hole), which give identical
contributions. For simplicity, we rewrite the Drude weight in terms of
only the $L-1$ string, identified as the particle string, with $\eta =
\eta_{L-1}$ and $\sigma=\sigma_{L-1}$ resulting in
%\begin{widetext}
\begin{equation}
\label{eq:zotosForm}
D= -\frac{J\sin\gamma}{2\pi \gamma\beta} \sigma\int_{-\infty}^\infty d\theta \frac{\left(\partial_\theta \ln \eta\right)^2}{\partial_\beta \ln \eta }
\frac{\eta \left(\partial_{\beta h} \ln \eta \right)^2}{ (1+\eta)^2}.
\end{equation}%\end{widetext}
In the infinite temperature limit, the above formula can be expanded
to leading order in inverse temperature
\cite{benz_finite_2005,urichuk_spin_2019}. The expansion is consistent with results obtained by an explicit construction of the conserved charges~\cite{prosen_families_2013}
%In the infinite temperature limit the Y-system can be expanded to leading order in inverse-temperature~\cite{benz_finite_2005,urichuk_spin_2018}, likewise the dressing relations can be solved directly from GHD~\cite{collura_analytic_2018}, and the quasi-local charges have been constructed at infinite temperatures~\cite{prosen_families_2013}. These approaches yield the exhausted Prosen bound
\begin{equation}
\label{Prosen}
16T D(\infty) = \frac{J^2
\sin^2\gamma}{\sin^2\left(\frac{\pi}{m}\right)}\left(1-
\frac{m}{2\pi} \sin\left(\frac{2 \pi}{m}\right) \right),
\end{equation} 
which, interestingly, is fractal in the anisotropy $\gamma = \frac{n
\pi}{m}$. This is in contrast to the $T=0$ case, where the Drude weight
can be obtained simply from the Kohn formula, Eq.~\eqref{Kohn},
because only the curvature of the ground state energy level
contributes. The continuous result is \cite{shastry_twisted_1990}
%In contrast with the case of $T=0$, where the Drude weight can be determined directly from Bethe ansatz~\cite{zvyagin_theory_1990,zvyagin_momentum_1991,shastry_twisted_1990} and found to be smooth in anisotropy
\begin{equation}
\label{D0}
D(0) = \frac{v_0K}{4 \pi},
\end{equation}
where the Luttinger parameter $K=\frac{\pi}{\pi-\gamma}$ and spinon
velocity $v_0=\frac{J\pi \sin\gamma}{2\gamma}$ are determined from the
Bethe ansatz. We note that both the infinite temperature result
\eqref{Prosen} and the zero temperature result \eqref{D0} are 
symmetric under $\gamma \to \pi-\gamma$ ($\Delta\to -\Delta$). This
is, however, not true for finite temperatures and the regimes of
positive and negative $\Delta$ both have to be analyzed.

A numerical evaluation of Zotos' formula~\eq{zotosForm} suggests that
for $T>0$ the zero field Drude weight is always fractal similar to the
infinite temperature case and that the leading low-temperature
correction for $0<\Delta<1$ scales as $T^{2K-2}$
\cite{zotos_finite_1999}, while the leading correction for $-1 < \Delta \leq 0$ scales as $T^2$. The low-temperature behavior has also been
investigated by a combined bosonization and memory matrix
approach~\cite{sirker_conservation_2011,sirker_diffusion_2009,sirker_transport_2020},
which predicts that at finite temperatures a ballistic and a diffusive
channel coexist. In this approach, Umklapp scattering is responsible
for partially melting the Drude peak at finite temperatures with the
difference going into the regular part of the conductance. However,
the temperature dependence of the Drude weight cannot easily be
obtained because it is related to the overlap of the current operator
with the quasi-local charges. The representation of the latter in
field theory is still an open issue.

Here, analytical results for the low-temperature Drude weight will be
obtained directly from Zotos' formula, supplemented by a field theory
approach for the leading temperature correction with integer
exponent. A combined auxiliary function~\cite{klumper_spin-1/2_1998}
and TBA approach~\cite{takahashi_thermodynamics_1999} will be
taken. Our calculation makes use of the relationship between the
$\eta$ functions and the quantum transfer matrix method in terms of
the Y-system and T-systems defined
in~Ref.~\cite{kuniba_t-systems_2011} via the fusion hierarchy, taking
advantage of the simplest parts of both approaches.

Let us first rewrite Eq.~\eqref{eq:zotosForm} in a form which is more
amenable to a low-temperature asymptotic analysis. In the following we set $J=1$.
 %For brevity, we will refer to the the second last string appearing\eqref{eq:zotosForm} as the `particle string' and the final string as the `hole string'. The order of the strings is based on the construction and conventions outlined in Ref.~\cite{kuniba_continued_1998}. The particle and hole strings are related to each other as $\rho_L = \rho_{L-1}^h=:\bar{\rho}_{L-1}$, which makes it possible to formulate Zotos' formula in terms of the particle string alone.
 The integrand in Eq.~\eqref{eq:zotosForm} is simplified by noting that
%$\partial_\theta \ln \eta = \dr{e}'
% \beta$, $\partial_\beta \ln \eta = \dr{e}$, and 
$\partial_{\beta h} \ln \eta = \frac{m}{2}$ and identifying an
effective velocity 
\begin{equation}
\label{eff_vel}
v=-\frac{\sin\gamma}{\beta \gamma} \frac{\partial_\theta \ln \eta}{\partial_\beta \ln \eta}
\end{equation} 
leading to 
\begin{equation}
%D &=\frac{\sin(\gamma)}{4 \pi \gamma \beta} \frac{2}{\sigma_{L-1}} \int d\theta \left(\frac{m}{2}\right)^2 \frac{(\dr{e}' \beta)^2}{\dr{e}} \frac{\eta}{(1+\eta)^2},\\
\label{eq:zotosSimplification}
D= -\frac{ m^2 \sigma}{4 \pi }\int_0^\infty d\theta \,v\, \partial_\theta \vartheta
%\frac{\partial_\theta\eta}{(1+\eta)^2}
\end{equation}
with the Fermi weight $\vartheta(\theta)= (1+\eta(\theta))^{-1}$. 
An integration by parts yields
\begin{equation}
\label{ZotosFinal}
D =- \frac{m^2 \sigma}{4 \pi}\left( \vartheta(\infty) v_0-\int_0^\infty d\theta\, \vartheta \,\partial_\theta v\right) ,
\end{equation}
with $v(\infty) = v_0$ and $v(0)=0$. The velocity changes rapidly
around $\theta=0$, changing from the left to the right moving spinon
velocity. This implies that the low-temperature Drude weight is
dominated by the region $\theta \sim 0$. In the following, we are
setting up the formalism to carry out an asymptotic analysis of this
region.

\section{Non-linear Integral Equations}
\label{NLIE}
A particularly useful approach to determine the asymptotics of
thermodynamic quantities is based on a set of non-linear integral
equations (NLIE). For the XXZ chain, they were originally derived
in~Ref.~\cite{klumper_thermodynamics_1993} and again in
Ref.~\cite{destri_unified_1995}. All thermodynamic quantities can be
expressed in terms of auxiliary functions $\aux$, $\bar{\aux}$, and
associated functions $\lna = \ln(1+\aux)$, $\lnb =
\ln(1+\bar{\aux})$, which are related to each other by
\begin{eqnarray}
\label{eq:NLIEdef}
\ln \aux &= -e_0 \beta + \kappa \star \lna - \kappa^{2\ii} \star \lnb, \,\nonumber \\
\ln \bar{\aux} &= -e_0 \beta + \kappa \star \lnb - \kappa^{-2\ii} \star \lna \, ,
\end{eqnarray}
with $g^{\ii c} = g(\theta+\ii c)$. Here $(\kappa^{\ii c} \star f)(\theta) = \int_{-\infty}^\infty d\mu \,\kappa
(\theta- \mu +\ii c) f(\mu)$ denotes a convolution with $c$ as a real
number. The kernel is given by 
\begin{equation}
\label{kappa}
\kappa(\theta) =
\int \frac{dk}{2\pi} \frac{\sinh\left((p_0-2)k\right) e^{\ii k\theta}}{2
\cosh(k)\sinh\left((p_0-1)k\right)}
\end{equation} 
with $p_0 = \frac{\pi}{\gamma}$, and $e_0 =
\frac{v_0}{\cosh\left(\frac{\pi}{2}\theta\right)}$. The main benefit of these NLIE is that the two auxiliary
functions contain all the thermodynamic information of the model,
which is in contrast with the large number of functions appearing
in the usual TBA formalism.

At low-temperatures, the auxiliary functions rapidly change around
$\theta \sim \shL =\frac{2}{\pi}\ln(v_0 \beta) $. By introducing the
shift $\theta \to \theta + \shL$ the temperature dependence will enter
the rapidity directly~\cite{aufgebauer_complete_2010}. After shifting the
arguments in the NLIEs the contributions from the negative and positive 
parts of the real axis may be separately considered. For Eq.~\eqref{eq:NLIEdef} we find, in particular,
\begin{widetext}\begin{eqnarray}
\label{NLIE2}
\ln &\aux &= - e_0 \beta + \int^\infty_0 d\mu\left[ \kappa(\theta-\mu)\lna(\mu)
-  \kappa^{2\ii}(\theta-\mu) \lnb(\mu)+  \kappa(\theta+ \mu) \lna(-\mu)
- \kappa^{2\ii}(\theta+ \mu) \lnb(-\mu)\right],\nonumber \\
\ln &\aux^\shL &\sim - 2\e^{-\theta \frac{\pi}{2}} + \int^\infty_{0} d\mu\left[ \kappa^\shL(\theta-\mu)\lna(\mu)+\kappa^{\shL}(\theta+ \mu)\lna(-\mu)
- \kappa^{\shL+2\ii}(\theta-\mu)\lnb(\mu) -  \kappa^{\shL+2\ii}(\theta+ \mu)\lnb(-\mu) \right]\nonumber\\
&&= -2 \e^{-\theta \frac{\pi}{2}} + \int^\infty_{-\shL} d\mu\left[ \kappa(\theta-\mu)\lna_+(\mu)
- \kappa^{2\ii}(\theta-\mu) \lnb_+(\mu)+  \kappa^{2\shL}(\theta + \mu) \lna_-(\mu)
- \kappa^{2\shL+2\ii}(\theta+ \mu) \lnb_-(\mu)\right]\nonumber\\
&&= - 2\e^{-\theta \frac{\pi}{2}} + \int^\infty_{-\shL} d\mu\left( \kappa(\theta-\mu)\lna_+(\mu)
- \kappa^{2\ii}(\theta-\mu) \lnb_+(\mu)\right) +\psi(\theta+2\shL),
\end{eqnarray}\end{widetext}
with $\lna_\pm(\mu) = \lna(\pm \mu \pm \shL)$. On the second line, we
substituted $\theta \to \theta + \shL$ resulting in $e_0 = \frac{v_0}{\cosh\frac{\pi}{2}\theta} \to \frac{2}{\beta} \e^{-\frac{\pi}{2}\theta}$, and on the third line $\mu \to
\mu + \shL$. We introduce the function $\psi(\theta+2\shL)$ on
the fourth line, which contains sub-leading temperature
contributions. 

As a preparation for calculating the low-temperature Drude weight, we
will first rederive the low-temperature asymptotics of the free energy
$f(T)$. Our calculations for the Fermi weight
$\vartheta(T)$ in Eq.~\eqref{ZotosFinal} will proceed
along similar lines. At low-temperatures, the auxiliary functions along the right/ left of the real axis are related to the free
energy as \cite{aufgebauer_complete_2010} 
\begin{eqnarray}
\label{free}
f&=& e_0 - \frac{T}{4} \int_{-\infty}^\infty d\theta \frac{\ln| A \bar{A}|}{\cosh\left(\frac{\pi}{2}\theta\right)} \nonumber\\
 &=& e_0 - \frac{T^2}{2 v_0}\int_{-\shL} d\theta \frac{\ln |A_+ A_- \bar{A}_+ \bar{A}_-|}{\e^{\pi\theta/2} } + O(T^4) \nonumber\\
 &\sim& e_0 - \frac{ T^2}{ v_0}\int_{-\shL} d\theta\, \e^{-\pi\theta/2}\ln |A_+ \bar{A}_+ |  ,
\end{eqnarray}
where $e_0$ is the ground state energy. The final line follows by
identifying $\lna_\pm = \lnb_\mp$. This free energy relation can be
used to construct a useful identity. First, we note that the
asymptotic limits of the auxiliary functions are
\begin{eqnarray}
\ln \aux_+(\infty)&=&\ln \bar{\aux}_+(\infty) = 0,\nonumber \\ \lna_+(\infty) &=& \lnb_+(\infty) = \ln 2 \, .
\end{eqnarray}
This allows us to derive the following identity via the so-called `dilog trick'
\begin{equation}
\frac{\pi^2}{3}=\int_{-\shL}^\infty d\theta \left( [\ln \aux_+]' \lna_+ - \ln\aux_+[\lna_+]'\right) + \text{c.c.} \, 
\end{equation}
On the r.h.s. we can now insert the NLIEs~\eqref{eq:NLIEdef}, \eqref{NLIE2},
which can be rewritten in terms of the free energy \eqref{free}
as
\begin{eqnarray}
\label{eq:dilogTrick}
\frac{\pi^2}{6}&=&  
%\int_{-\shL}^\infty d\theta\left[\pi\frac{ \ln|A_+ \bar{A}_+|}{\e^{\theta \frac{\pi}{2}}}+(\psi^{2\shL})' \lna_+ + (\bar{\psi}^{2\shL})'\lnb_+\right] \nonumber\\ 
\int_{-\shL}^\infty d\theta \left[\pi\e^{-\theta \frac{\pi}{2}} \ln|A_+ \bar{A}_+|
%&+&\int_{-\shL}^\infty d\theta\left[  (\psi^{2\shL})' \lna_+ + (\bar{\psi}^{2\shL})'\lnb_+\right] \nn \\
+\left( (\psi^{2\shL})' \lna_+ + \text{c.c.} \right)\right] \nn \\
&+&\lna_+ \star \partial_\theta \kappa \star \lna_+ -\lnb_+ \star \partial_\theta \kappa^{2\ii} \star \lna_+ \nonumber \\
&+&\lnb_+ \star \partial_\theta \kappa \star \lnb_+ -\lna_+ \star \partial_\theta \kappa^{-2\ii} \star \lnb_+ \, ,\nonumber\\
\frac{\pi^2}{6}&= &\frac{\pi v_0}{T^2}( e_0 - f )  + \int_{-\shL}^\infty d\theta \left[ (\psi^{2\shL})' \lna_+ + \text{c.c.}\right] \, ,
\end{eqnarray}
with the understanding that $\lna_+ \star \partial_\theta \kappa \star \lna_+ = \int_{-\shL}^\infty d\theta\int_{-\shL}^\infty d\mu \lna_+(\theta) \partial_\theta (\kappa(\theta- \mu)) \lna_+(\mu)$. These terms vanish due to $\lna_+ \star \partial_\theta \kappa \star \lna_+ =-\lna_+ \star \partial_\mu \kappa \star \lna_+$, whereas $\lnb_+ \star \partial_\theta \kappa^{2\ii} \star \lna_+ = - \lnb_+ \star \partial_\mu \kappa^{-2\ii} \star \lna_+$  exactly cancel each other.

The leading order temperature dependence of $(\psi^{2\shL})' =
\partial_\theta \psi(\theta+2\shL)$ is obtained by expanding the
kernel $\kappa(\theta-\mu)$ in Eq.~\eqref{kappa}. Temperature
corrections with integer exponents correspond to the pole at $\frac{
\ii \pi(2q-1)}{2}$, with $q\in\mathbb{Z}_{>0}$. The dominant
low-temperature correction to the Drude weight for $0<\Delta < 1$
comes from the pole at $\frac{\ii \pi}{p_0-1}$. In the following, we
will concentrate on the contribution from this pole and will return to
the temperature corrections with integer exponents in
Sec.~\ref{Numerics}. We can expand around the pole at $\frac{\ii
\pi}{p_0-1}$ straightforwardly by use of the kernel definition,
resulting in
\begin{equation}
\label{eq:kAsymp}
\kappa(\theta-\mu+2\shL) \sim \frac{(K-1)}{2 } \tan(\pi K) \e^{-\ka(\theta-\mu)} \e^{-2\ka \shL}
\end{equation}
with $\ka = \frac{\pi \gamma}{\pi - \gamma}= \pi(K-1)$. We note that if $\gamma = \frac{\pi}{3}$ ($\ka = \frac{\pi}{2}$) then $\tan(\pi K) \to \pm \infty$. This expansion
may then be used to obtain the $\psi(\theta+2\shL)$ functions which
are then inserted back into~\eqref{eq:dilogTrick}, yielding an integral
relation between the leading order temperature contribution of these
functions and the low-temperature free energy 
%The $\psi(\theta+2\shL)$ functions yield, to leading order in temperature, with $\ka = \frac{\pi \gamma}{\pi - \gamma}= \pi(K-1)$
%\begin{widetext}
\begin{eqnarray}
\label{eq:freeAuxiliary}
 \psi^{2\shL}&=&\ka \e^{-\ii \ka}\tan(\pi K)\left(\frac{T}{v_0}\right)^{4K-4} \nonumber \\
&\times&\left[  \int_{-\shL}^\infty \frac{d\mu}{2 \pi}\, \e^{-\ka(\theta+\mu-\ii)}\lna_- - \text{c.c.}\right],\nonumber\\
 f-e_0 &=&-\frac{\pi T^2}{6 v_0}+\frac{v_0}{2} (K-1)^2 \tan(\pi K)\left(\frac{T}{v_0}\right)^{4K-2}\nonumber \\
&\times & 
\left[\int_{-\shL}^\infty d\mu \, \e^{-\ka (\mu-\ii)}\lna_- - \text{c.c.}\right]^2 \\
&=&-\frac{\pi T^2}{6 v_0}+\frac{v_0}{2}(K-1)^2 \tan(\pi K)  \left(\frac{T}{v_0}\right)^{4K-2 }I^2. \nonumber
\end{eqnarray}
%\end{widetext}
In the second line, the $\psi^{2\shL}=\psi(\theta+2\shL)$ functions
are inserted into~\eq{dilogTrick} with $\lna_\pm = \lnb_\mp$, which
results in two sign changes. 

At low-temperatures, the remaining unknown integral $I$ can be determined
by comparing the result above with Ref.~\cite{lukyanov_low_1998}. In
this work, a comparison of a zero temperature, finite field
Wiener-Hopf calculation with field theory leads to 
\begin{eqnarray}
\label{eq:freeFieldTheory}
f &= & e_0 - \frac{\pi T^2}{6v_0} - \frac{\sin(2\pi K) }{v_0}(2T\gC)^2 \left[ \frac{\sqrt{\pi} T}{v_0}\right]^{4K-4} ,\nonumber\\
\gC&=&\frac{\Gamma^2\left( K\right)\Gamma\left(1- 2K\right)}{\Gamma^2\left(1-K\right)
}\left[ \frac{\Gamma\left(1+ \frac{1}{2K-2}\right)}{\Gamma\left(1+ \frac{K}{2K-2}\right)}\right]^{2K-2}.
\end{eqnarray}
Comparing Eqs.~\eq{freeAuxiliary} and~\eq{freeFieldTheory} we find
that the integral $I$ is given by
\begin{eqnarray}
%\frac{\sqrt{\tan(\pi(K-1))} I(0)}{\sqrt{2}  v_0^{2K-\frac{3}{4}}(K-1)^{-1} } &=&\pm i \frac{2\gC \sqrt{\sin(2\pi K)}}{\sqrt{v_0}\left(v_0/\sqrt{\pi}\right)^{\frac{2\ka}{\pi}}},\nonumber\\
\label{eq:intIdentity}
I = \pm \ii \frac{4 \pi^{K-1}}{(K-1)}\gC\cos(\pi K).
\end{eqnarray}
We will see later that the same integral appears in the
calculation of the low-temperature asymptotics of the Drude weight and
that the negative sign is the proper choice for the Drude weight
to decrease with increasing temperature. In addition to this
physical argument, the negative sign is confirmed by numerical
computations.

% Nov 17

\section{Low-temperature Fermi weight}
\label{Fermi}
Having understood how to obtain the low-temperature asymptotics of the
free energy using the NLIE, we are now ready to return to
Eq.~\eqref{ZotosFinal}. To calculate the low-temperature Drude weight,
the Fermi weight $\vartheta$ must be determined for $\theta\sim
0$. The relationship between the $\eta$ functions and the quantum
transfer
matrix~\cite{benz_finite_2005,kuniba_t-systems_2011,urichuk_spin_2019}
yield an identity that determines the Fermi weight of the particle
string in terms of the $q$ and $\phi$ functions (see
App.~\ref{fermiIdentity} for a detailed derivation
of~\eqref{eq:fermiWeightEval} and App.~\ref{ratioFunction} for
the integral relations used for evaluating the $\phi/qq$ functions)
%\begin{widetext}
\begin{eqnarray}
\label{eq:fermiWeightEval}
m\vartheta(\theta) &=&\sum_{j=1}^{\bar{\mu}} \frac{\phi^{\ii (s+2j)}(\theta)}{q^{\ii (s+2j-1)}(\theta)q^{\ii (s+2j+1)}(\theta)}, \\
\phi(\theta) &=& \left(\sinh\left[\frac{\gamma(\theta+\ii u+\ii)}{2} \right]\sinh\left[\frac{\gamma(\theta-\ii u-\ii)}{2} \right]\right)^{N/2}, \nn \\
q(\theta)&=&\prod_{ \{\lambda_j\}}\sinh\left[\frac{\gamma (\theta-\lambda_j)}{2}\right],\nn
\end{eqnarray}%\end{widetext}
with $\lambda_j\in \{\lambda_j\}$ being Bethe
roots~\cite{kuniba_continued_1998}, $N$ being the Trotter number, $u= -\frac{ J \sin{\gamma}}{T \gamma N}$, and 
\begin{equation}
\label{eq:TSshift}
s = 2m - 2\bar{\mu} + \frac{(-1)^{1+\alpha}}{n}-p_0-1
\end{equation}
stemming from the string integers. Here $\bar\mu$ is the length of the
hole string and $\alpha$ is the length of the continued fraction for
$p_0$, see App.~\ref{AppA} for details. The $\phi/qq$ functions can be
related to the auxiliary functions $\aux,\bar{\aux}$ by revising the
derivation of the NLIE
\cite{klumper_thermodynamics_1993,klumper_conformal_1992}, see also App.~\ref{ratioFunction}.

We first analyze the case $p_0=\pi/\gamma=m \in
\mathbb{Z}$ in which case the hole string has length $\bar{\mu}=1$, $\alpha=1$, and the Fermi
weight simplifies to $\vartheta = \frac{1}{p_0}\frac{\phi^{\ii
p_0}}{q^{\ii (p_0+1)} q^{\ii(p_0-1)}}$. In terms of the auxiliary functions,
we can then express the Fermi weight as
\begin{equation}
\label{eq:integralRelation}
\log \left( \frac{\phi^{\ii p_0}}{q^{\ii (p_0+1)}
	q^{\ii(p_0-1)}} \right) = \omega^\ii \star \lnb- \omega^{-\ii} \star \lna,
\end{equation}
where the kernel is given by 
\begin{equation}
\label{omega_kernel}
\omega(\theta) =  \int \frac{dk}{2\pi}
\frac{\e^{\ii k\theta}}{2\sinh((p_0 -1)k)}
\end{equation}
and $\lna, \lnb$ are defined in Eq.~\eq{NLIEdef}.

In the scaling limit, $\theta \to \theta +
\frac{2}{\pi}\log(v_0\beta)$~\cite{aufgebauer_complete_2010}, the
kernels may be expanded in temperature. This procedure is analogous to
the expansion of the kernels $\kappa(\theta)$ that leads to
Eq.~\eqref{eq:kAsymp}. The dominant contributions for $0<\Delta<1$
stem from the poles at $k=0$ and $k=\frac{\ii \pi}{p_0-1}$ taking the
form
\begin{equation}
%\omega(x) = \frac{1}{2\pi}& \int d\theta \frac{\e^{i kx}}{2\sinh((p_0 -1)k)}\\
\label{eq:omegaAsymp}
\omega(\theta+\shL) \sim \frac{\ii\ka}{4\pi} - \frac{\ii \ka}{2 \pi} \e^{-\ka \theta} \e^{-\ka \shL}.
\end{equation}
In contrast to Eq.~\eqref{eq:kAsymp}, there is now also a zeroth order
contribution. For $\vartheta(\theta)$ with $\theta \ll \shL$ we can
now evaluate the r.h.s. of Eq.~\eqref{eq:integralRelation}
\begin{widetext}
\begin{eqnarray}
\label{wkernel}
&& \omega^\ii \star \lnb -\omega^{-\ii} \star \lna=\int_{-\shL}^\infty d\mu\left[  \omega^{\ii-\shL}(\theta-\mu)\lnb_+-\omega^{-\shL-\ii}(\theta-\mu) \lna_+ +\text{c.c.}\right] \nonumber\\ %-\omega^{\shL-i}(\theta+\mu) \lna_-+\omega^{\shL+i}(\theta+\mu) \lnb_-\right] \nonumber\\
&&=\int_{-\shL}^\infty\!\! d\mu\left[ \omega^{\shL+\ii}(-\theta+\mu) \lna_+ - \omega^{\shL-\ii}(-\theta+\mu)\lnb_+ +\text{ c.c.}\right] %, \nonumber\\%+\omega^{\shL-i}(\theta+\mu) \lna_--\omega^{\shL+i}(\theta+\mu) \lnb_-\right] \nonumber\\
=\int_{-\shL}^\infty\!\! d\mu\left[ (\omega^{\shL+\ii}(-\theta+\mu)+\omega^{\shL+\ii}(\theta+\mu)) \lna_+ +\text{ c.c.}\right]\nonumber\\%+ (\omega^{\shL-i}(-\theta+\mu) +\omega^{\shL-\ii}(\theta+\mu))\lnb_+ \right] \nonumber\\
&&\sim V_0 +\frac{\ii\ka\left(\e^{\ka\theta}+\e^{-\ka \theta}\right)}{2\pi}\left(\frac{T}{v_0}\right)^{2K-2} \int_{-\shL}^\infty d\mu\left[\e^{-\ka (\mu-\ii)} \lnb_+ -\e^{-\ka(\mu+\ii)} \lna_+ \right] \nonumber\\
%&\omega^{-i}& \star \lna - \omega^i \star \lnb
&&\sim V_0+ \ii (K-1) I \cosh\left(\ka\theta\right) \left(\frac{T}{v_0}\right)^{2K-2}=V_0+ 4 \gC \cos(\pi K) \cosh\left(\ka\theta\right) \left(\frac{\sqrt{\pi}T}{v_0}\right)^{2K-2}.
% minus sign correct?
\end{eqnarray}\end{widetext}
Here $V_0$ is the $T=0$ contribution which we will determine
later. The identity $\omega(\theta) = -\omega(-\theta) +\text{const}$
is used on the second line and then both $\lna_\pm = \lnb_\mp$
and Eq.~\eq{omegaAsymp} are inserted.
The integral $I$ is exactly the same one as in the low-temperature
expansion of the free energy, see Eq.~\eq{intIdentity}. From
Eq.~\eqref{wkernel} we therefore find for $\theta \ll \shL$
\begin{eqnarray}
&&\frac{\phi^{\ii p_0}}{q^{\ii(p_0-1)}q^{\ii(p_0 + 1)}}= \exp\left(\omega^\ii \star \lnb-\omega^{-\ii} \star \lna \right),\nonumber\\
\label{eq:ratioFunctionExp}
&\sim& \e^{\left[V_0+ 4\gC \cosh(\ka\theta)\cos(\pi K)\left(\frac{\sqrt{\pi}T}{v_0}\right)^{2K-2}\right]}.
%&\sim\frac{e^{V_0}}{1+4\gC(K) \cosh(\alpha\theta)\cos(\alpha)\left(\frac{\sqrt{\pi}T}{v_0}\right)^{2K-2}}.
\end{eqnarray} 
For the considered case $p_0=\pi/\gamma=m \in \mathbb{Z}$, where $\bar{\mu}=1$, the small rapidity Fermi weight is found to be
\begin{equation}
\label{FermiWeight}
\vartheta % =& \frac{1}{m} \frac{\phi}{q^i q^{-i}} \\b
\sim \frac{\e^{V_0}/m}{1-4\gC \cosh(\ka\theta)\cos(\pi K)\left(\frac{\sqrt{\pi}T}{v_0}\right)^{2K-2}}.
\end{equation}
The leading finite-temperature dependence for $0<\Delta<1$ is thus
accounted for, but the $T=0$ contribution must be determined. This is
possible because the zero-temperature Drude weight is
known~\cite{shastry_twisted_1990}, providing a second identity. So
returning to Zotos' formula~\eqref{ZotosFinal} with
$\vartheta(0)=\vartheta_0=\e^{V_0}/m$ when $T=0$ it follows that
\begin{equation}
D_0 = \frac{v_0 K}{4\pi} = -\frac{m^2 v_0 \sigma}{4 \pi} \left( \vartheta(\infty) - \vartheta_0\right).
\end{equation}
$\vartheta(\infty) = \frac{\bar{\mu}}{m}$ is determined by 
looking at the large rapidity asymptotics of Eq.~\eq{fermiWeightEval}. 
The zero temperature Fermi weight $\vartheta_0$ is then found to be
\begin{equation}
\label{eq:fermiValue}
\vartheta_0 = \frac{\left(\sigma K + \bar{\mu}  m\right)}{m^2} =\frac{\sigma +(m-n) \bar{\mu} }{m(m-n)} \, .
\end{equation}
For the simple roots of unity, $p_0=\pi/\gamma=m$, this expression
reduces to $\vartheta_0 = \frac{1}{m-1}=K-1$.
%Move the general anisotropy to later % General anisotropies result in $\e^{V_0} = m\frac{\vartheta_0}{\bar{\mu}-k}$. The integer $k$ is defined by $m = n \ell +u$ and $\bar{\mu}= k \ell+c$ with $0\leq c<k$ and $0< u<n$ and for $\bar\mu=1$ that $k:=0$, $c=1$. 

\section{Low-Temperature Drude Weight}
\label{Asymp1}
Now that we have determined the low-temperature Fermi weight
\eqref{FermiWeight} for the simple roots of unity, 
we are in a position to evaluate
Eq.~\eqref{ZotosFinal}. First, the low-temperature Fermi
weight can be used to determine the effective velocity $v$ to leading
order
\begin{eqnarray}
\label{eq:velEff}
v &=&\frac{\sin(\gamma)}{\gamma T}\frac{\partial_\theta \ln (\eta)}{\partial_T \ln (\eta)}= \frac{\sin(\gamma)}{\gamma T}\frac{\partial_\theta \ln (\vartheta)}{\partial_T \ln (\vartheta)},\nonumber\\
 &\sim& v_0 \tanh(\ka \theta).
\end{eqnarray}

Defining $B=4 \gC
\cos(\pi K)\left(\frac{\sqrt{\pi}}{v_0}\right)^{2K-2}$ and
noting that $\sigma=1$ when $\gamma = \frac{\pi}{m}$, Zotos' formula
then reads to leading order
%\begin{widetext}
\begin{eqnarray}
&&D=\frac{m^2 }{2 }\left[\int^\infty_0 \frac{d\theta}{2\pi} \frac{v_0 \ka \vartheta_0/\cosh^{2}(\ka \theta)}{1-B\cosh(\ka \theta)T^{2K-2}}-\frac{v_0\vartheta(\infty)}{2\pi}\right] \nonumber\\
&& \sim \int^\infty_0 \frac{d\theta}{4\pi}\frac{v_0 m^2\ka \vartheta_0}{\cosh^{2}(\ka \theta)}\left[1 + {B T^{2K-2}}{\cosh(\ka \theta)}\right]-\frac{v_0 m}{4\pi }. \nonumber \\
&& 
\end{eqnarray}
%\end{widetext}
These integrals can be evaluated noting that $\int_0^\infty d\theta \frac{\ka}{ \cosh^2(\ka \theta) } =1$, and $\int_0^\infty d\theta \frac{\ka}{ \cosh(\ka \theta) } =\frac{\pi}{2}$. The Drude weight is then given by
\begin{equation}
D(T) \sim \frac{m^2v_0}{4\pi} \left( \vartheta_0 - \frac{1}{m} + \frac{\pi}{2} \vartheta_0 B T^{2K-2} \right).
\end{equation}
We can rewrite the result for $\gamma = \pi/m$ in terms of the Luttinger parameter
$K$ using $\vartheta_0 = K-1$ and $m = \frac{K}{K-1}$. Our final
result for the leading temperature asymptotics of the Drude weight of
the XXZ chain at the simple roots of unity $\gamma=\pi/m$ then reads 
\begin{equation}
\label{final1}
D \overset{\bar{\mu}=1}{=} D_0 + \frac{v_0 K^2}{2(K-1)}\gC \cos(\pi K)\left(\frac{\sqrt{\pi}}{v_0}T\right)^{2K-2}.
% &=& D_0 - \frac{m v_0}{2}\Theta(\gamma)\gC(K) \cos(\alpha)\left(\frac{\sqrt{\pi}}{v_0}\right)^{2K-2}.
\end{equation}
%For anisotropies $\pi/\gamma \in \mathbb{Z}$ then $\Theta(\gamma) = \e^{V_0} = m\vartheta_0$. %For more general anisotropies the zeroth temperature contribution instead results in $\Theta(\gamma) = m\frac{\vartheta_0}{\bar{\mu}-k}\frac{\sin\left(\frac{\pi K}{m}\right)}{\sin(\pi K)}$ with the integer $k$ defined from the hole string length $\bar{\mu} = k \ell + c$ with $m = n \ell +u$ for $0\leq c<k$ and $0<u<n$.
Note that the temperature dependence agrees with the one found first
numerically by Zotos \cite{zotos_finite_1999} and that the correction
is overall negative. We note that Eq.~\ref{final1} is only valid at
the discrete anisotropies $K= \frac{m}{m-1}$, $m=2, 3 \dots$ with
$\lim_{K\to 2} \gC \cos(\pi K) = 0$ and $\lim_{K\to 3/2} \gC \cos(\pi
K) = -\frac{\sqrt{\pi}}{48}$, noting that the divergence of $\gC$ for
$K \to \frac{3}{2}$ is cancelled by $\cos(\pi K)$ leading to a finite
value. This correction, due to Umklapp scattering, vanishes at the
free fermion point ($K=2$) and the leading temperature correction
$\sim T^2$ stems from band curvature terms. These integer scaling
dimensions are related to those poles of $\omega(\theta)$ which are
independent of the anisotropy $\gamma$. We discuss this point further in
section~\ref{Numerics} where we extend the asymptotics to the regime
$-1<\Delta\leq 0$. Furthermore, we have checked that the prefactor of
the $T^{2K-2}$ correction does agree with a numerical evaluation of
Zotos' formula \eqref{eq:zotosForm}, see also
Sec.~\ref{Numerics}. Finally, we observe that the result only depends
on the Luttinger parameter $K$ and the spinon velocity $v_0$. The
Drude weight of the XXZ model at anisotropies $\Delta=\cos(\pi/m)$
thus fits into the usual Luttinger liquid universality.
%Eq.~\eqref{final1} should thus be, at least in
%principle, also derivable using field-theoretical methods.

% Nov 14

\section{General Anisotropies}
\label{Asymp2}
We will now extend our results from the simple roots of unity case
$\gamma=\pi/m$ to general anisotropies $\gamma=n\pi/m$. We will
continue to concentrate on the $T^{2K-2}$ temperature correction which
is the dominant one for $0<\Delta<1$. We expect a result which is more
complicated than \eqref{final1}, since we know that the Drude weight
at infinite temperature has a fractal structure
\cite{prosen_families_2013} and expect that this remains true at all
finite temperatures. On the technical side, the calculation becomes
more complicated because the sum in~\eq{fermiWeightEval} now contains
multiple terms. In order to determine the sum, the imaginary shift
must be carefully considered. The integral expression
\eqref{eq:integralRelation} is valid only for $\Im(\theta) \in
(2-p_0,p_0-2)$, whereas $\frac{\phi^{\ii p_0}}{q^{\ii p_0-\ii}q^{\ii
p_0+\ii}}(\theta)$ is analytic in the region $\Im (\theta) \in (1-p_0,
p_0-1)$ and the narrower strip $\Im(\theta) \in (-1,1)$. In order to
carry out a similar calculation as that done for the simple root of
unity case, the analytic region of the integral relation
Eq.~\eq{integralRelation} must be considered, which is determined by
the analytic region of the kernel $\omega(\theta)$ in
Eq.~\eqref{omega_kernel}. The argument of $\phi/qq$ is cyclic under
shifts of $2\ii p_0$, which is useful for keeping the argument in the
analytic region of the integral relation. Even with this cyclic
relation, however, some shifts fall outside of the analytic
region. These cases, as we will show below, can be treated in one of
two ways: first by an identification between the ratio functions and
the largest eigenvalue of the quantum transfer matrix, and secondly
through analytic continuation of the $\omega$-kernel.

When $\Im(\theta)=0$ modulo $2p_0$, the ratio functions in
Eq.~\eqref{eq:fermiWeightEval} may develop poles along the integration
axis. To resolve this, the pair of ratios
$\frac{\phi^\ii}{q^{2\ii}q}+\frac{\phi^{-\ii}}{q^{-2\ii}q}$, which
always appear together, must be considered as a single object. The
largest eigenvalue of the quantum transfer matrix is given by \cite{kuniba_continued_1998} $\Lambda(\theta)=
\frac{\phi^\ii q^{-2\ii}}{q}+\frac{\phi^{-\ii} q^{2\ii}}{q}$ which yields
the relation
\begin{equation}
 \frac{\Lambda}{q^{-2\ii}q^{2\ii}}= \frac{\phi^{\ii}}{q^{2\ii}q}+\frac{\phi^{-\ii}}{q^{-2\ii}q}.
\end{equation}
A similar integral expression as Eq.~\eq{integralRelation} may be
determined for the pair of ratio functions (see
App.~\ref{ratioFunction} for a derivation)
\begin{equation}
	\label{eq:freeEnergyTrick} \ln \frac{\Lambda}{q^{-2\ii}q^{2\ii}}=\omega^{\ii(2-p_0)}\star\lnb- \omega^{-\ii(2-p_0)}\star \lna.  
\end{equation} 
There is a sign flip in the leading temperature term relative to the simple root of unity case Eq.~\eqref{eq:ratioFunctionExp}, explicitly
\begin{equation}
\label{eq:lambdaTrick}
\frac{\Lambda}{q^{-2\ii}q^{2\ii}} \sim \exp\left(V_0 -B \cosh(\ka \theta)T^{2K-2}\right).
\end{equation}
This relation can then be used to evaluate string length $\bar{\mu}=2$ cases
analogously to the simple root of unity case. When $\bar{\mu}=2$ and
$\cos(\gamma)>0$, the anisotropy reads $\gamma =
\frac{m-1}{2} \frac{\pi}{m}$, with odd m, so $p_0=\frac{2m}{m-1}=[2,\frac{m-1}{2}]$ 
and the length of the continued fraction is given by $\alpha
=2$. Eq.~\eqref{eq:TSshift} then reduces to
\begin{equation}
s +2 p_0 \overset{\bar{\mu}=2}{=}2m - 3.  
\end{equation}
The sum in Eq.~\eqref{eq:fermiWeightEval} can now be evaluated using
the integral relation Eq.~\eqref{eq:freeEnergyTrick} and the cyclic
property $2m\ii = 2(m-1) p_0 \ii$ leading to
\begin{eqnarray}
\vartheta(\theta) &=&\frac{1}{m}\sum_{j=1}^{2} \frac{\phi^{\ii(2m-3+2j)}}{q^{\ii(2m-4+2j)}q^{\ii(2m-2+2j)}} \\
&=& \frac{1}{m} \left( \frac{\phi^{-\ii}}{q^{-2\ii}q}+\frac{\phi^{\ii}}{q^{2\ii}q}\right) % \nonumber\\
\sim \frac{1}{m}\frac{e^{V_0}}{1+B \cosh(\ka \theta)T^{2K-2}}. \nonumber
\end{eqnarray}
The zero temperature Drude weight is again used to
determine the value of $e^{V_0}/m$. When 
$\bar{\mu}=2$ the variable $\sigma=-1$, which results
in the correct sign for the Drude weight correction.

%{\color{red} Start of changes}
A second scenario occurs when the argument of (\ref{eq:integralRelation})
falls outside the analytical strip of the $\omega$-kernels appearing in the
integral relations. These terms may be treated by analytical continuation
of first the kernel and second the convolution integrals. The function \eqref{omega_kernel}
% $\omega(\theta) = \int_{-\infty}^\infty \frac{dk}{2\pi} \frac{\e^{\ii k\theta}}{2\sinh((p_0-1)k)}$ 
is defined and analytic in $ \Im(\theta) \in
(1-p_0 , p_0-1)$
with poles at $\pm \ii(p_0-1)$. This can be seen from the asymptotics of
the integrand
\begin{equation}
\frac{\e^{\ii k\theta}}{2\sinh((p_0-1)k)}%{\e^{(p_0-1)k}-\e^{(1-p_0)k}}
 \sim\begin{cases}
 \e^{\ii k\theta} \e^{(1-p_0)k}&  \, k\to \infty\\
-	\e^{\ii k\theta} \e^{(p_0-1) k}& \, k\to -\infty 
\end{cases} .
\end{equation}
%This indicates that the $\omega$-kernel converges in the region
%$1-p_0<\Im (\theta) <p_0-1$.
Hence, the integral expression Eq.~\eqref{eq:integralRelation} is surely analytic for
$ \Im(\theta) \in (2-p_0 , p_0-2)$. The analytic continuation outside this
strip is possible. Here we show how to analytically continue downwards into 
$\Im(\theta) \in (-p_0, 2-p_0)$ for which the convolution with the kernel
$\omega^{-\ii}$ is relevant. The analytic continuation upwards into $\Im(\theta)
\in (p_0-2,p_0)$ is done analogously which covers the entire periodicity strip $\Im(\theta)
\in (-p_0,p_0)$.
The analytic continuation of the kernel
is based on separating and explicitly integrating the leading asymptotics of
the integrand of the Fourier integral
%\begin{widetext}
\begin{eqnarray}
	\omega(\theta) &=& \int_{-\infty}^0 \frac{dk}{2\pi}
	\frac{\e^{\ii k\theta} }{\e^{(p_0-1)k}-\e^{(1-p_0)k}} \nonumber \\
&+&\int_0^\infty \frac{dk}{2\pi} \frac{\e^{\ii k\theta}
	e^{(2-2p_0)k}}{\e^{(p_0-1)k}-\e^{(1-p_0)k}} + \int_0^\infty \frac{dk}{2\pi} \e^{\ii k\theta} e^{(1-p_0)k}\nonumber \\ % \label{omegaa}\\
        &=&\dr{\omega}(\theta)+\frac1{2\pi}\frac \ii {\theta+\ii(p_0-1)}=:\omega_c(\theta)\label{omegab}
\end{eqnarray}
%\end{widetext}
Here the first two integrals converge in the region $\Im(\theta)
\in (3-3p_0 , p_0-1)$ where they define the analytic
$\omega_c(\theta)$ appearing in the last line of (\ref{omegab}), and
the third integral was evaluated explicitly giving the simple rational
function with pole at $-\ii(p_0-1)$. Repeating this argument shows
that $\omega(\theta)$ is a meromorphic function with poles at
arbitrary odd integer multiples of $\pm \ii(p_0-1)$.

Next we consider the analytic continuation of the convolution
\begin{equation}
  \omega\star\lna(\theta)=\int d\mu \, \omega(\theta- \mu)\lna(\mu)\,,
\end{equation}  
using the pole $-\ii(p_0-1)$ of $\omega$ with residue $\ii/2\pi$. Letting the
imaginary part of the argument $\theta$ move from above $-(p_0-1)$ to below enforces
the deformation of the $\mu$-integration contour into the lower half plane for
rendering the dependence on $\theta$ analytic. The deformed contour may then be
replaced by a straight $\mu$-contour and a closed contour surrounding
$\theta+\ii(p_0-1)$ in counter-clockwise manner. The convolution with standard
$\mu$-contour has to be evaluated with the expression (\ref{omegab}) for
$\omega(\theta-\mu)$, the closed contour integral yields the explicit term
$\lna(\theta+\ii(p_0-1))$ such that
\begin{equation}
  (\omega\star\lna)_c(\theta)=  \omega_c\star\lna(\theta)+\lna(\theta+\ii(p_0-1))\,,
\end{equation}  
meaning that the continuation of $\omega\star\lna$ is identical to the
convolution of the continued $\omega$ with $\lna$ plus the explicit term.
In $\Im(\theta) \in (-p_0, 2-p_0)$ we have the analytic continuation
\begin{eqnarray}
  &&\omega^{\ii} \star \lnb -(\omega^{-\ii}\star \lna)_c  \nn\\
  &&={\omega}^{\ii} \star
  \lnb -{\omega}^{-\ii}_c \star \lna - \lna^{\ii(p_0-2)}.
\end{eqnarray}
For the case of $\Im(\theta) \in (-p_0,2-p_0)$ thus only one single
additional explicit term appears. This explicit term can be determined
from Eq.~\eq{NLIEdef}, where it is negligible when $\Im(\theta) \in
(1-p_0,2-p_0)$, whereas for $\Im(\theta) \in (-p_0,1-p_0)$ it has a
very negative real part.  Explicitly, $\lna^{\ii (p_0 -2)}=\ln (1 +
\aux)^{\ii (p_0 -2)}$ is determined to leading order from $\ln
\aux^{\ii(p_0-2)}(\theta) \sim - \beta e_0(\theta+\ii(p_0-2))$, which
has very negative (positive) real part for $\Im(\theta)$ above (below)
$1-p_0$. To summarize: when $\Im(\theta) \in (1-p_0,p_0 -2)$ the ratio
function may be treated by using $\omega_c$ in
the integral expression, but when $\Im(\theta) \in (-p_0,1-p_0)$ the ratio
function vanishes.
%{\color{red} END}

Drude weights at arbitrary anisotropies $\gamma=n\pi/m$ are calculable
by using the considerations outlined above. The calculation requires
$\Im(\theta)$ to be classified in order to determine the ratio
function contribution. This can be carried out, for example, by fixing
the numerator $n$ of the anisotropy and determining $\bar{\mu}$ based
on the denominator $m$. Cases up to $n=7$ were explicitly calculated,
which led to a conjecture for the general form of the Drude weight
correction that was checked against low-temperature numerical results
and found to be consistent.

For another concrete example of this procedure, let us consider the
case $n=3$. There are two possible continued fractions for these
anisotropies: $p_0 = \pi/\gamma = \left[\frac{m-1}{3},3\right]$, and $p_0 =
\left[ \frac{m-2}{3},1,2\right]$. Anisotropies corresponding to 
$p_0 = \left[\frac{m-1}{3},3 \right]$ yield shifts according to Eq.~\eq{TSshift}:
\begin{equation}
s = \frac{m-4}{3} - 2\bar{\mu} ,\, \text{ mod}\left(2p_0\right),
\end{equation} 
with $p_0 = m/3$, where $m=6 \ell +1$ or $m=6 \ell+4$. These two
scenarios may be separately treated by identifying exceptional terms,
where a pair of ratio functions are evaluated by the eigenvalue trick
or there is a vanishing ratio function. Notably, for $m=7$ the result
for $\bar{\mu}=2$ applies, but when $m\geq 7$ other shifts appear in
addition to those terms that require the eigenvalue trick to
evaluate. In this case terms in~\eqref{eq:fermiWeightEval} with overall shift $0$ or $\pm 1$
lead to a vanishing contribution from the ratio function or a
pair of ratio functions that must be evaluated by the eigenvalue
trick, respectively. These exceptional terms occur for
\begin{equation}
%2j \leq \frac{4-m}{3},\,\text{or} \,
j = \begin{cases}
\ell+1 \,& \, \text{if } \, m=6\ell+4\, \text{ and } \, \bar{\mu} = 2\ell+1,\\
\ell, \ell+1 \,& \, \text{if } \, m=6\ell+1\, \text{ and }\, \bar{\mu} = 2\ell.
\end{cases}
\end{equation}
These special cases must be treated separately from the other
shifts. For $m=6\ell+1$ the eigenvalue trick is used to evaluate the
paired set of shifts.  In order to compute the Fermi weight, these
shifts are given by $2j - 2\bar{\mu} +\frac{5 m - 4}{3}$, where the
last string length is given by $\bar{\mu}=2 \ell$. 
To stay in the analytic strip of~\eqref{eq:integralRelation} the $2\ii p_0$ periodicity of the functions may be exploited. By adding or subtracting shifts of $2\ii p_0$ the remaining shifts can be made to fit in the analytic domain as
\begin{eqnarray}
2 j -4 \ell +\frac{6\ell-3}{3} +p_0 =
2 j -\frac{2}{3},\nn \\
2 j -4 \ell +\frac{6\ell-3}{3} -p_0 =
2 j- 4 \ell -\frac{4}{3}.
\end{eqnarray}
The first set is in the analytic strip when $j = 1, \dots \ell-1$ and the second for $j = \ell+2, \dots 2 \ell$. The index $j$ may be rewritten in the second shift as $j' = j - \ell -1$, so the sum goes from $j' = 1, \dots \ell-1$. These considerations along with the special terms now allow evaluation of the sum
\begin{widetext}\begin{eqnarray}
\vartheta&=&\frac{m-4}{3m} \e^{V_0} + \left[\sum_{j=1}^{\frac{m-7}{6}}\left(\e^{\ii\ka\left(2j-\frac{2}{3}\right)}+\e^{\ii\ka\left(2j-\frac{m-1}{3}+\frac{2}{3}\right)}\right)-1\right]\frac{\e^{V_0}}{m}B \cosh(\ka \theta) T^{2K-2} ,\\
&=&\vartheta_0+ \left[\frac{ \sin\left(\frac{\ka (m-6)}{3}\right)-\sin\left(\frac{\ka}{3}\right) }{\sin(\ka)}-1\right]\frac{3 \vartheta_0}{(m-4)}B \cosh(\ka \theta) T^{2K-2}
=\vartheta_0- \frac{3\vartheta_0\sin\left(\frac{\ka}{3}\right)}{(m-4)\sin(\ka)}B \cosh(\ka \theta) T^{2K-2}.\nonumber %= \vartheta_0-\frac{\vartheta_0(0)\sin\left(\frac{\pi K}{m}\right)}{(\bar{\mu}-1)\sin(\pi K)}B \cosh(\alpha \theta) T^{2K-2}.
\end{eqnarray}\end{widetext}
At this point, the Drude weight is evaluated in 
the same way as for the simple root of unity case. This type of calculation 
can be carried out for any anisotropy, but is generally quite tedious. 

A further simplification is possible by noting $\sin(\dr{K})=-\sin(\pi
K)$ and $\sin\left(\frac{\dr{K}}{3}\right)=-\sin(\pi K/m)$ for
$n=3$. Based on these considerations, we find that the Drude weight
correction, $\Delta D =D_0-D=a(\Delta)T^{2K-2}$, which is the leading
term for $0<\Delta<1$, is given by
\begin{eqnarray}
\label{eq:drudeFinal}
\Delta D&=& \frac{\sin\left(\frac{\pi
K}{m}\right)}{\tan(\pi K)} \frac{v_0 m K }{2}\gC \left[ \frac{
\sqrt{\pi}T}{v_0}\right]^{2K-2},\nonumber\\ 
\Delta D &\overset{\bar{\mu}=1}{=}& -\cos(\pi K)\frac{v_0 K^2}{2
(K-1)} \gC \left[ \frac{ \sqrt{\pi}T}{v_0}\right]^{2K-2}\!\!\! .
\end{eqnarray}
On the second line, the general formula from the first line is specialized to 
the simple root of unity case $\gamma=\pi/m$ ($\bar\mu=1$). At simple roots of unity, the result can be expressed entirely by the
Luttinger liquid parameter $K$ and the spinon velocity $v_0$. This is,
however, not the case for $\bar\mu>1$, where a factor $m
\sin\left(\frac{\pi K}{m}\right)$ appears. This suggests that our general
result is outside of standard Luttinger liquid theory. It is exactly
this direct dependence on the integer $m$ which makes the prefactor
fractal as a function of anisotropy. This is similar to the result obtained
for the infinite temperature Drude weight in Eq.~\eqref{Prosen}.
\begin{figure}
%\subfloat[a]{
\includegraphics[width=0.95\columnwidth]{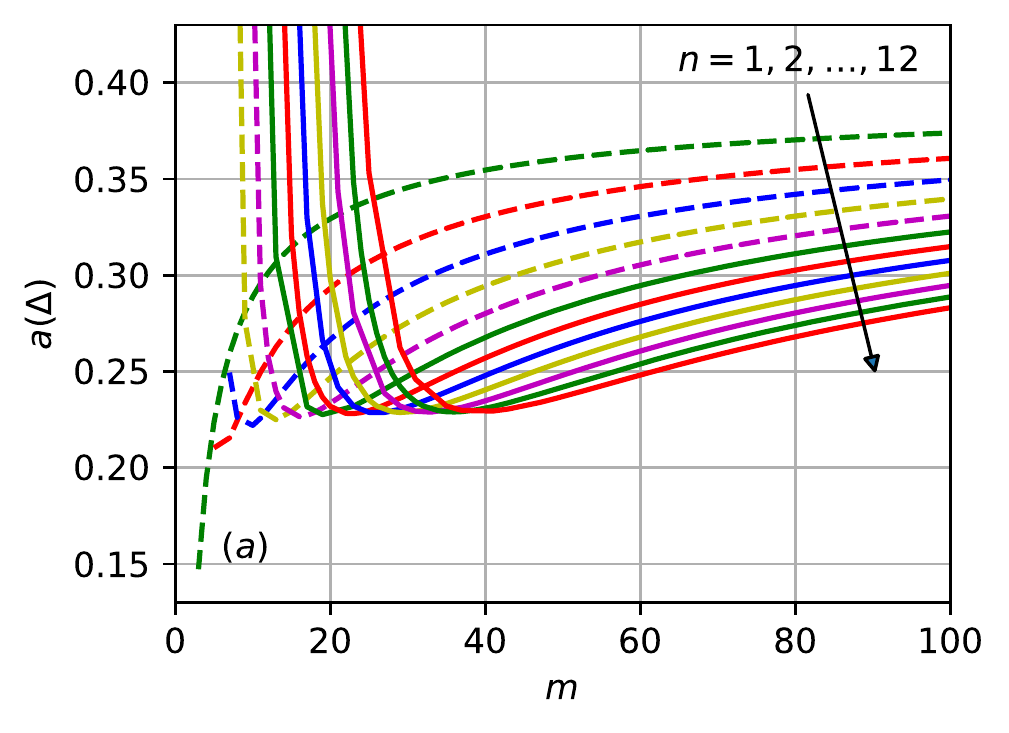}
%\subfloat[]
\includegraphics[width=0.95\columnwidth]{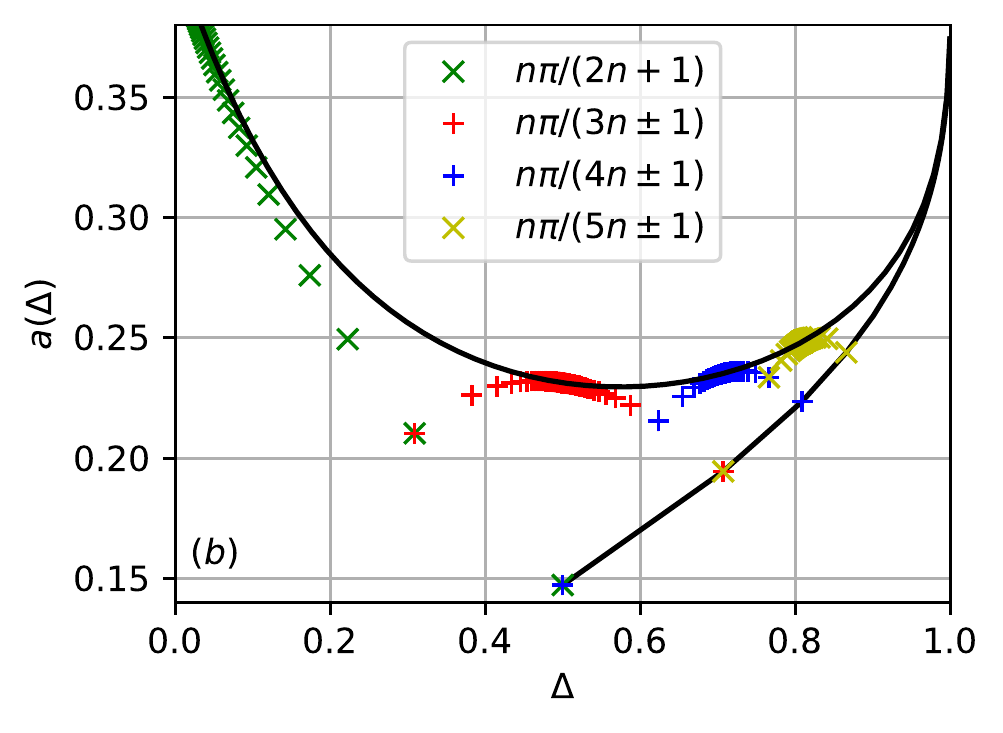} 
\caption{(a) For each fixed $n$, the function $a(\Delta)=a(\cos(n\pi/m))$ is smooth. 
(b) When ordered by anisotropy $\Delta$, however, the prefactor
$a(\Delta)$ is fractal. Here the upper solid line denotes the envelope
function obtained from the result \eqref{eq:drudeFinal} for
$\bar\mu>1$ in the formal limit $m\to\infty$ with $K$ fixed. The lower
solid line is the result \eqref{eq:drudeFinal} for $\bar\mu=1$
analytically continued for $K\leq 3/2$ ($\Delta\geq 1/2$).}
\label{Fig1}
\end{figure}
In Fig.~\ref{Fig1}, the prefactor $a(\Delta)$ is plotted for various
fixed $n$. While each curve for a given $n$ is a smooth function of
$m$, the dependence on anisotropy $\Delta=\cos(\gamma)$ is clearly
fractal. Our results thus provide further evidence that $D(\Delta)$ is
analytic in $\Delta$ only for $T=0$, while it is a fractal for $T>0$.

\begin{figure}
\includegraphics[width=0.95\columnwidth]{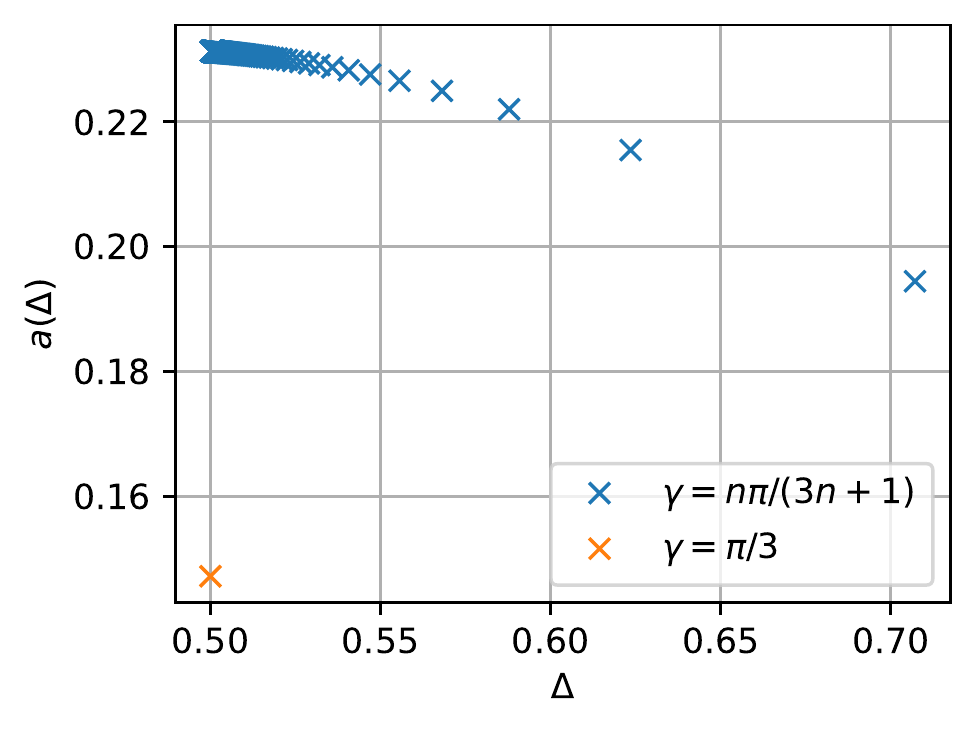}
\caption{Prefactor $a(\Delta)$ for constant string length $\bar{\mu}=3$ approaching $\gamma=\frac{\pi}{3}$. 
There is a clear jump at $\gamma=\frac{\pi}{3}$ which exemplifies the fractal
character of the Drude weight at low temperatures.}
\label{Fig2}
\end{figure}

This structure may be further analyzed by considering limiting cases of
Eq.~\eqref{eq:drudeFinal}. A particularly notable curiosity is the observation
that for anisotropies $\gamma = \frac{n \pi}{n+1}$ $(K= n+1)$ the correction factor $a(\Delta) =0$. 
For other anisotropies $\gamma=\frac{n \pi}{m}$ with fixed $n$, the Drude weight correction term $a(\Delta)$
also falls onto continuous curves for fixed string length
$\bar\mu$. Along these curves, limiting values of the Drude weight
correction may be determined for anisotropies intersecting with these
curves. The possible limits that can be carried out are $m \to
\infty$, for fixed $n$, and $\frac{1}{p_0}=\frac{\gamma}{\pi} \to c$ for fixed
$\bar{\mu}>1$, which are accomplished by treating the denominator $m$ as a function of the numerator $n$ such that the string length is kept fixed. 
We note that $\gC$ diverges like $m$ as $\gamma \to \pi/3$ and otherwise converges to a
finite constant. From the trigonometric functions in
Eq.~\eqref{eq:drudeFinal} it is then obvious that $\lim_{\Delta \to
\frac{1}{2}} a(\Delta) \neq a\left(\frac{1}{2}\right)$, which is what
we mean when calling the Drude weight fractal. As an example, the
curve with constant string length $\bar\mu=3$ approaching
$\gamma=\pi/3$ clearly demonstrates this fractal property as can be
seen in Fig.~\ref{Fig2}.

We note that the prefactor $a(\Delta)$ in Eq.~\eqref{eq:drudeFinal} is
always finite. There are no divergencies. Furthermore, we note that
our analytical low-temperature asymptotics do not apply at the
isotropic point $\Delta=1$ ($K=1$). Here the ordering of the
temperature corrections collapses: terms such as $T^{r(4K-4)}$ with
$r$ integer which are next-leading for $K>1$ all show the same
temperature scaling as the $T^{2K-2}$ term in the limit $K\to 1$. A
numerical evaluation of Zotos' formula for $m\to\infty$ indicates that
$D(T>0)=0$ at the isotropic point, which is consistent with the
quasilocal charges becoming non-local for $\Delta \to
1$~\cite{prosen_exact_2011,prosen_families_2013,pereira_exactly_2014}. This
vanishing of the Drude weight at the isotropic point has been used to
argue in support of superdiffusive transport at
$\Delta=1$~\cite{ilievski_superdiffusion_2018,ljubotina_spin_2017}.

\section{Integer exponents and numerical results}
\label{Numerics}
So far, we have concentrated on the pole of the kernel
$\omega(\theta)$ leading to the $T^{2K-2}$ temperature correction. In
addition, there are poles which are independent of anisotropy and lead
to temperature corrections $T^{2r}$ with $r\in\mathbb{N}$ as well as
higher order corrections $\sim T^{r(2K-2)}$, $r=2,3,\cdots$. In
principle, one could try to extend the asymptotic analysis of the TBA
equations discussed in the previous sections to obtain results beyond
the $T^{2K-2}$ temperature correction. Having seen how technically
demanding it is to obtain just the $T^{2K-2}$ asymptotics, we will
however instead argue that other corrections---in particular the $T^2$
correction which is dominant for $-1<\Delta\leq 0$---can be obtained
in a standard field theoretical calculation and are, in fact, already
known, see
Refs.~\cite{sirker_conservation_2011,sirker_diffusion_2009}.

\begin{figure}
\includegraphics[width=0.95\columnwidth]{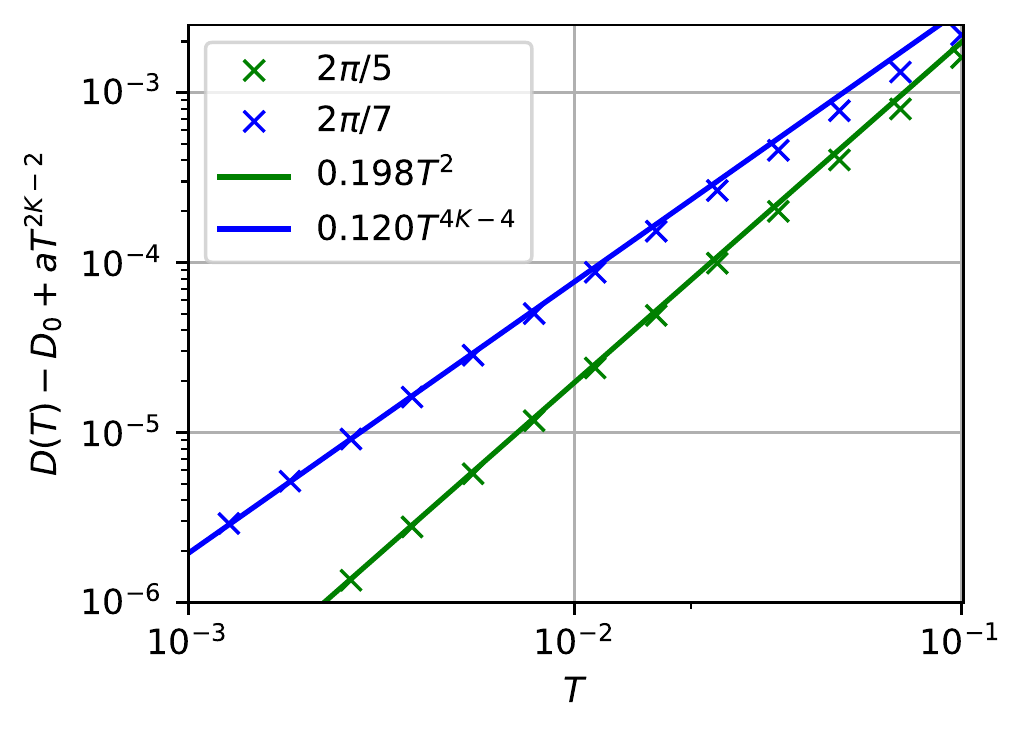}
\caption{Numerical results for $D(T)$ based on Eq.~\eqref{eq:zotosForm} 
with the zero-temperature Drude weight $D(0)$ and the leading
temperature correction $aT^{2K-2}$ subtracted for
$\Delta=\cos(2\pi/5)\approx 0.31$ and $\Delta=\cos(2\pi/7)\approx
0.62$. For $\gamma=2\pi/5$ the next-leading correction scales as $T^2$
while it scales as $T^{4K-4}$ with $K=7/5$ for $\gamma=2\pi/7$.}
\label{Fig_FT1}
\end{figure}

We start by numerically studying the next-leading temperature
corrections in the two regimes $0<\Delta<1/2$ and $1/2<\Delta<1$. As shown
in Fig.~\ref{Fig_FT1}, a power law in temperature can be clearly
identified with an exponent which is different in the two regimes.
In a field theoretical description of the XXZ chain, we obtain the
standard free boson theory in the scaling limit with perturbations
stemming from band curvature and from Umklapp scattering. From the
scaling dimensions of these operators, we expect that band curvature
leads to temperature-dependent corrections of the Drude weight $\sim
T^{2r}$, with $r$ a positive integer, while Umklapp scattering results
in terms scaling as $T^{r(2K-2)}$. The numerical findings in
Fig.~\ref{Fig_FT1} are in agreement with these expectations. For
$0<\Delta <1/2$ ($3/2<K<2$) the next-leading correction scales as $T^2$
while the temperature dependence changes to $T^{4K-4}$ for
$1/2<\Delta<1$ ($1<K<3/2$).

In Refs.~\cite{sirker_conservation_2011,sirker_diffusion_2009}, a standard bosonization
approach was used to calculate the spin conductivity
$\sigma(q,\omega)$ as a function of momentum $q$ and frequency
$\omega$. Band curvature and Umklapp scattering were taken into
account by a perturbative calculation of the retarded self-energy
$\Pi^{\textrm{ret}}(q,\omega)$ leading to
\begin{equation}
\label{SelfE}
\Pi^{\textrm{ret}}(q,\omega)\approx -2\ii \gamma\omega + b\omega^2+cv_0^2 q^2 \, .
\end{equation}
The problem with this result is that the imaginary part of the
self-energy, characterized by the relaxation rate $\gamma$, will lead
to a complete decay of the current-current correlation function
$\langle J_0(t) J_0(0)\rangle/N \sim \exp(-2\gamma t)$; there is no
Drude weight. The reason is that the integrable structure of the
microscopic model---in particular, the existence of quasi-local
conserved charges---is not reflected in the standard field theoretical
approach. We want to argue here that this issue only affects the
imaginary part of the self energy and that the real part is not
affected by the quasi-local charges and does contain information about
other temperature corrections to the Drude weight. In particular,
perturbations of the free boson model due to band curvature do not
relax the current---independent of whether or not local conservation
laws exist---and can be treated in a standard perturbative
manner. Setting the relaxation rate $\gamma=0$---which would
correspond to purely ballistic transport---the Drude weight in this
field theory approach is given by (see Eq.~(2.52) in
Ref.~\cite{sirker_conservation_2011})
\begin{equation}
\label{FT1}
D(T) =\frac{Kv_0}{4\pi(1+b(T))}
\end{equation}
with 
\begin{equation}
\label{FT2}
b=(Y_2-Y_3)T^{4K-4}+Y_4 T^2 \, .
\end{equation}
Here the $T^{4K-4}$ term is obtained in second order perturbation
theory in Umklapp scattering while the $T^2$ term is obtained in first
order in band curvature. Due to the integrability of the model, the
amplitudes $Y_2,\, Y_3$, and $Y_4$---which are functions of $K$, $v_0$
only---can be determined. They are given in
Ref.~\cite{sirker_conservation_2011} and we will reproduce them for
convenience in App.~\ref{FT_App}. The hypothesis that \eqref{FT2}
contains parts of the additional temperature corrections to the Drude
weight is supported by a comparison of Eq.~\eqref{FT1} with the Bethe
ansatz calculations based on spinons and anti-spinons in
Ref.~\cite{benz_finite_2005}. The latter approach also appears to
assume that spin transport is purely ballistic at all
temperatures. The results of the two approaches at low temperatures
are in excellent agreement, see Fig.~2 in
Ref.~\cite{sirker_conservation_2011}.

The partial decay of the Drude weight is
caused by Umklapp scattering which turns two right movers into left
movers and vice versa and is therefore able to relax the part of the
current which is not protected by the quasi-local conserved
charges. While the proper field theoretical treatment of the
quasi-local charges is not yet known, the scaling dimension and the
prefactor of the leading temperature correction in
Eq.~\eqref{eq:drudeFinal} show that this term corresponds to a
correction which is first order in Umklapp scattering. In second order
in Umklapp scattering, which leads to a $T^{4K-4}$ temperature
correction, there are two contributions: a contribution which does not
change the current and is contained in Eqs.~(\ref{FT1},\ref{FT2}), and a
contribution which does change the current and which we have not
determined here.

We therefore conjecture that the Drude weight of the XXZ chain for
$-1<\Delta <1$ at low temperatures is asymptotically given by
\begin{equation}
\label{FT3}
D(T) = D_0 -aT^{2K-2} -b_1T^2 -b_2 T^{4K-4}
%D(T) = \frac{Kv_0}{4\pi}[1 - (Y_2-Y_3+c)T^{4K-4} - Y_4 T^2] - a T^{2K-2} 
\end{equation}
with $D_0=Kv_0/4\pi$ and $a$ being the prefactor in
Eq.~\eqref{eq:drudeFinal} which does depend, in general, on $m$ and
$K$. Here we have defined $b_1= D_0Y_4$ and $b_2=D_0(Y_2-Y_3+c)$ where
$Y_2,\, Y_3,\, Y_4$ are the amplitudes given in
Eq.~\eqref{FT_App1}. The term $\sim cT^{4K-4}$ is the current relaxing
contribution which we expect to occur in second order in Umklapp
scattering with an unknown amplitude $c$. The amplitude $Y_4$ diverges
whenever the exponent of one of the higher order Umklapp contributions
becomes equal to $2$, i.e. when $r(4K-4)=2$ which is equivalent to
$K=(2r+1)/(2r)$. These divergencies have to be cancelled by the
amplitudes of those parts of the higher Umklapp terms which are not
current relaxing. We will show this explicitly for $K=3/2$ below. For
the amplitude $c$ of the current relaxing part, on the other hand, we
expect no divergencies but---similar to the $T^{2K-2}$ term---a
fractal dependence on $n,m$. Although we do not know the amplitude of
the $T^{4K-4}$ temperature correction completely, we keep the known
part because the divergence of $Y_4$ would otherwise make the formula
\eqref{FT3} unusable for $K\sim 3/2$ ($\Delta\sim 1/2$). We stress
again that the $T^2$ term is the dominant temperature correction for
$-1<\Delta\leq 0$ and including this term is thus essential to obtain
an asymptotic result which is valid in the entire regime
$-1<\Delta<1$.

\begin{figure}
\includegraphics[width=0.95\columnwidth]{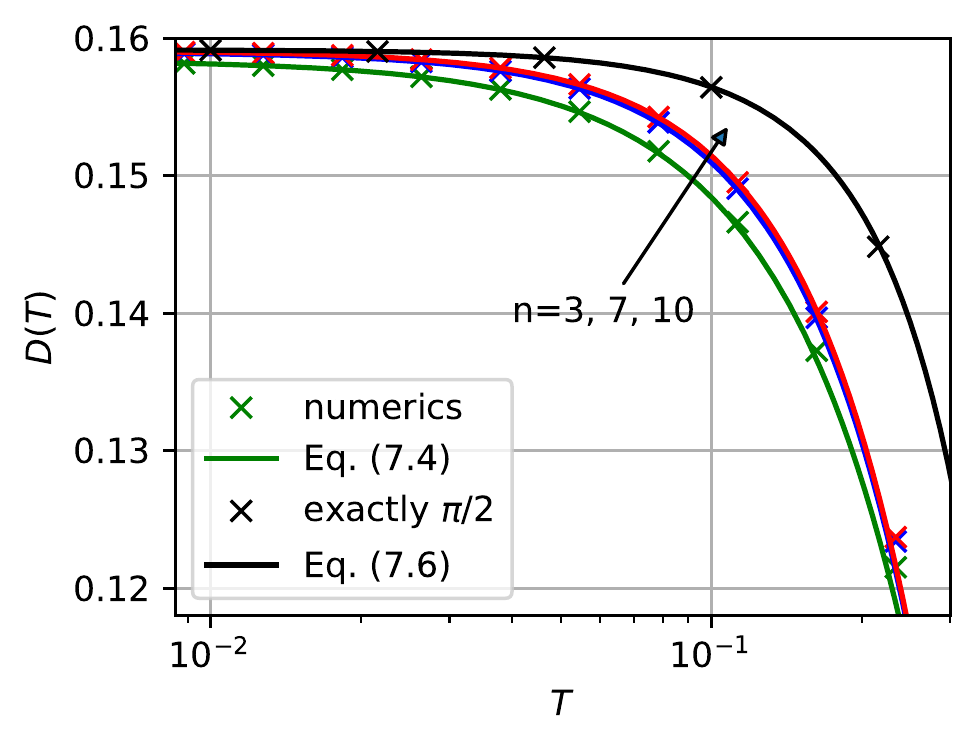}
\caption{Numerical results for $D(T)$ compared to the analytical result 
\eqref{FT4} at the free fermion point and the asymptotics \eqref{FT3} 
for anisotropies near the free fermion point,
$\gamma=n\pi/(2n+1)$. The asymptotics
\eqref{FT3} works well for temperatures $T\lesssim 0.2$.}
\label{Fig_FT2}
\end{figure}

Our hypothesis \eqref{FT3} can be checked directly for the free
fermion point $\Delta=0$. Here the current operator $J_0$ is conserved
and the Drude weight is simply given by a static expectation value
\begin{equation}
\label{fF}
D=\frac{\langle J_0^2\rangle}{2NT} = \frac{1}{8\pi T}\int_{-\pi}^\pi\frac{\sin^2(k)\, dk}{\cosh\left(\frac{\varepsilon_k}{T}\right)+1}
\end{equation}
with $\varepsilon_k=-\cos k$. The leading temperature dependence can
be obtained by an asymptotic evaluation of the integral and one finds
\begin{equation}
\label{FT4}
D(T) = \frac{1}{2\pi} -\frac{\pi}{12} T^2 - \frac{7\pi^3}{240} T^4 -\mathcal{O}(T^6) \, .
\end{equation}
In Eq.~\eqref{FT3}, the amplitudes $a$ and $Y_2-Y_3$ vanish for
$\Delta=0$ as can be checked explicitly but is also obvious because
Umklapp scattering, which is responsible for the temperature
corrections with non-integer exponents, is not present at the free
fermion point. Finally, $b_1\to \pi/12$ for $\Delta\to 0$ showing that
the leading temperature corrections are consistent with the free
fermion result. We therefore expect that $D(T)$ near the free fermion point is described over a
fairly large temperature range by Eq.~\eqref{FT3}. This is indeed the
case, see Fig.~\ref{Fig_FT2}.
Note that keeping only the leading $T^{2K-2}$ term, in contrast,
describes the data for $0<\Delta \ll 1$ well only at extremely low
temperatures.

\begin{figure}
\includegraphics[width=0.95\columnwidth]{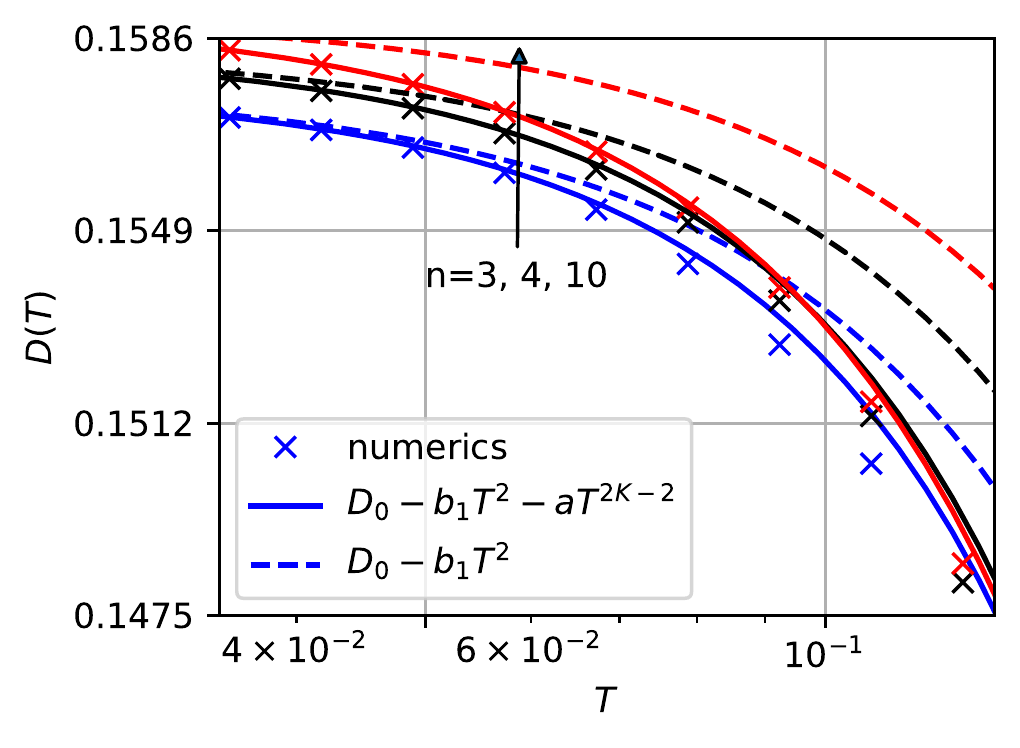}
\caption{Numerical results for $D(T)$ at $\gamma=n\pi/(2n-1)$ with $n=3,4,10$ 
compared to the analytical result \eqref{FT3} with and without the
next-leading $T^{2K-2}$ term.}
\label{Fig_negDelta}
\end{figure}

Let us now also check the regime $-1<\Delta\leq 0$ where the $T^2$
term is the leading temperature correction. For $-1/2<\Delta<0$ the
$T^{2K-2}$ term is next-leading while a $T^4$ term will be
next-leading for $-1<\Delta<-1/2$. As shown in
Fig.~\ref{Fig_negDelta}, Eq.~\eqref{FT3} is consistent with the
numerical data for the Drude weight at anisotropies
$\gamma=n\pi/(2n-1)$ with the next-leading $T^{2K-2}$ correction being
important to describe the data well up to temperatures $T\lesssim
0.1$.

The fractal structure of the Drude weight at finite temperatures is
very pronounced near the free fermion point. Right at the free fermion
point we have found the low-temperature asymptotics
\eqref{FT4}. Approaching the free fermion point by anisotropies
$\gamma=n\pi/(2n+1)$ with $n\to\infty$, on the other hand, we find
that the prefactor $a$ in Eq.~\eqref{FT3} does not vanish but rather
takes the limiting value $a\to \pi^2/24$. The leading temperature
dependence is therefore given by
\begin{equation}
\label{FT5}
\lim_{n\to\infty}D_{\gamma=n\pi/(2n+1)}(T) = \frac{1}{2\pi} -\left(\frac{\pi^2}{24}+\frac{\pi}{12}\right)T^2 +\mathcal{O}(T^4)
\end{equation}
in contrast to the result right at the free fermion point,
Eq.~\eqref{FT4}.

\begin{figure}
\includegraphics[width=0.95\columnwidth]{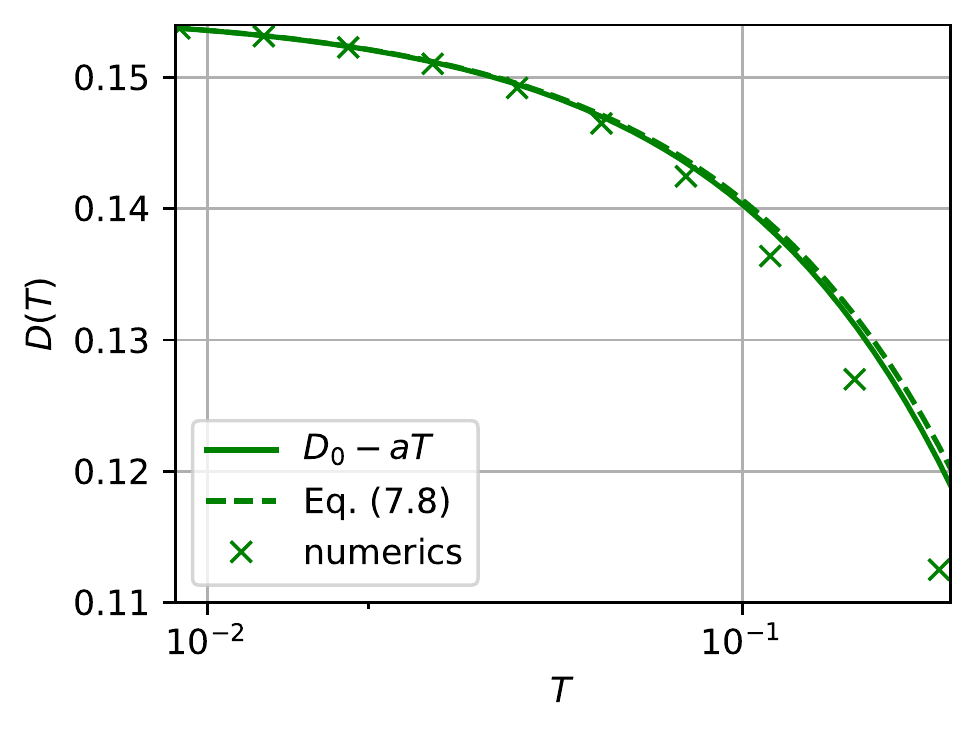}
\caption{Numerical results for $D(T)$ at $\Delta=1/2$ 
compared to the analytical result \eqref{FT3} with and without the
known part of the next-leading term.}
\label{Fig_FT3}
\end{figure}

Next, we consider the special case $\Delta=1/2$ ($K=3/2$). At this
point, both the next-leading contribution from Umklapp scattering and
the band curvature term scale as $T^2$ and both have divergent
amplitudes. These divergencies, however, cancel and Eq.~\eqref{FT3} yields
\begin{eqnarray}
&D&_{\gamma=\pi/3}(T)=\frac{9\sqrt{3}}{32\pi}-\frac{3\pi}{64}T+\frac{1}{12\sqrt{3}\pi}T^2\ln T +cT^2 \nn \\
\label{FT6}
&-& \frac{142+24\tilde\gamma+48\ln 2+60\ln3-21\zeta(3)}{288\sqrt{3}\pi}T^2 \, .
\end{eqnarray}
which includes the current relaxing second order Umklapp contribution
with unknown amplitude $c$. As shown in Fig.~\ref{Fig_FT3}, this
result without the $cT^2$ term is in good agreement with the numerical
data.
We note that in contrast to anisotropies near the free fermion point,
keeping the known parts of the next-leading terms does not increase
the temperature range over which the asymptotics agrees well with the
numerical results. We want to stress again, however, that by keeping
the known part of the $T^{4K-4}$ correction, the result \eqref{FT3} is
not plagued by divergencies for anisotropies $0\leq \Delta\leq 1/2$
either.

\begin{figure}
\includegraphics[width=0.95\columnwidth]{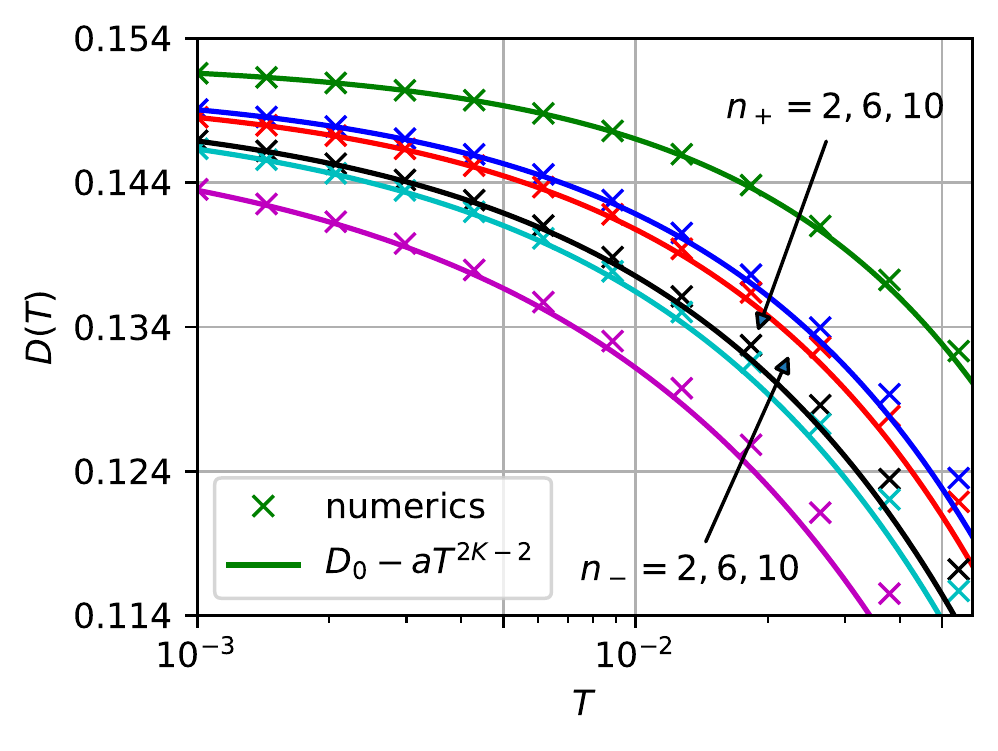}
\caption{Numerical results for $D(T)$ compared to the leading terms in the asymptotics 
for anisotropies $\gamma=n_+\pi/(4n_+ + 1)$ (circles and solid lines) and
$\gamma = n_- \pi /(4n_--1)$.}
\label{Fig_FT4}
\end{figure}

Finally, we consider anisotropies when approaching other simple roots of unity.
For the approach towards $\gamma =\pi/4$ shown in Fig.~\ref{Fig_FT4}, the leading temperature
correction describes the data well in the shown temperature regime.
Generally speaking, the derived asymptotics holds over a smaller and
smaller low-temperature range the closer we get to the isotropic
point. This is a consequence of all the different temperature
corrections $T^{r(2K-2)}$ with $r\in\mathbb{N}$ collapsing for $K\to
1$. Directly at the isotropic point, our result is not applicable.

\section{Conclusions}
\label{Concl}
We have obtained an analytic result for the leading low-temperature
asymptotics of the Drude weight of the XXZ chain at anisotropies
$\Delta=\cos(\pi n/m)$. Previously, analytical results were only known
at zero and infinite temperatures. Similar to the infinite temperature
case, we find that the leading low-temperature correction for
$0<\Delta<1$ is a fractal as a function of anisotropy $\Delta$. From a
more technical perspective, the temperature dependence of the Drude
weight at low temperatures in this regime is a consequence of the
combined small rapidity behavior of the effective velocity and the
Fermi weight. Our analytical result agrees with numerical evaluations
of Zotos' formula and adding the $T^2$ correction known from a field
theory approach yields a good description of the Drude weight over a
finite low-temperature range for the entire regime $-1<\Delta<1$. The
exception is the isotropic limit, $\Delta\to 1$, where this range
shrinks to zero.

While the result for the simple roots of unity $\gamma=\pi/m$ can be
expressed entirely by the Luttinger parameter $K$ and the spin
velocity $v_0$, this is not the case for general anisotropies
$\gamma=\pi n/m$. In the latter case, the $T^{2K-2}$ Drude weight
correction contains a factor $m\sin(\pi K/m)$ which is responsible for
the fractal structure. This puts the latter result outside the usual
Luttinger liquid framework. How such a factor can emerge in a field
theoretical description of this integrable lattice model remains an
open question.

\acknowledgments
A.U., J.S., and A.K. acknowledge sypport by the German Research
Council (DFG) via the Research Unit FOR 2316. J.S. acknowledges support
by the Natural Sciences and Engineering Research Council (NSERC,
Canada). 

\appendix
\section{Takahashi-Suzuki Integers}
\label{AppA}
The Takahashi-Suzuki (TS) integers are described by a set of recursive relations that categorize
the Bethe strings by their length ($\mu_j$), parity ($v_j$), and the sign of the dressed energy ($\sigma_j$). Continued fractions of the anisotropy are given by $\pi/\gamma = [\nu_1, \dots, \nu_\alpha]$. With 
\begin{equation}
[\nu_1,\nu_2,\nu_3, \dots] = \nu_1 + \frac{1}{\nu_2+\frac{1}{\nu_3 + \dots}}
\end{equation} 
these give rise to the $m$-integers
\begin{equation}
m_j = \sum_{\ell=1}^j \nu_\ell.
\end{equation}
The $\nu_i$ integers that appear in the continued fractions yield relations for TS-integers
\begin{eqnarray}
y_j= y_{j-2}+ \nu_j y_{j-1},\, \, 1 \leq j \leq \alpha,\nn\\
y_{-1}=0, \, y_0 =1 , \, y_1 = \nu_1\, ,
\end{eqnarray}
as well as for the $p$-numbers
\begin{eqnarray}
p_j = p_{j-2} - \nu_{j-1}p_{j-1}, \, \text{ for }  2 \leq j \leq \alpha +1,\\
p_0 = \frac{\pi}{\gamma}, \, p_1=1, \,  \text{ and } p_{\alpha+1}=0\nn.
\end{eqnarray}
With the above numbers, the string lengths are then determined by
\begin{eqnarray}
\mu_j = y_{\ell-1} + (j-m_\ell) y_\ell , \, \text{ for } \, m_\ell \leq j < m_{\ell+1},\nn\\
m=y_\alpha = \mu_{m_{\alpha}-1}+\mu_{m_{\alpha}}.
\end{eqnarray}
The string parity is characterized by
\begin{eqnarray}
z_j= z_{j-2}+ \nu_j z_{j-1},\, \text{ for }\, 1 \leq j \leq \alpha\nn\\
z_{-1}=1, \, z_0 =0 , \, z_1 = 1.
\end{eqnarray}
The parity is given in terms of the string length and anisotropy $\gamma = \frac{\pi n}{m}$ as 
\begin{eqnarray}
v_j &=& (-1)^{\left\lfloor \frac{{(\mu_j -1) n}}{m}\right\rfloor}, \text{  if } j \neq m_1,\nn\\
v_{m_1} &=& -1.
\end{eqnarray}
Finally, the numbers $q_j$ are given by
\begin{eqnarray}
q_j &=& (p_i - (j-m_i) p_{i+1}) (-1)^{i}, \, \, \text{ for } m_{i} \leq j < m_{i+1},\nn\\
q_{m_{\alpha}}&=&-q_{m_{\alpha}-1},
\end{eqnarray}
which can be used to determine $\sigma_j =\text{sign}(q_j)$.

\section{Largest Eigenvalue}
\label{ratioFunction}
The largest eigenvalue may be written in terms of the auxiliary functions by using a method analogous to the one used to derive \eq{integralRelation} and the non-linear integral relations themselves. In order to be consistent with earlier definitions, the Fourier transformation (FT) is defined as $f(k) = \int d\theta \e^{-\ii k\theta}f(\theta)$. The largest eigenvalue of the quantum transfer matrix is related to the auxiliary functions as

\begin{eqnarray}
\Lambda &=&\frac{q^{2\ii}\phi^{-\ii}}{q}+\frac{q^{-2\ii}\phi^{\ii}}{q},\nonumber\\
\Lambda^\ii &=& \frac{q^{-\ii}\phi^{2\ii}}{q^\ii}\left(1+\frac{q^{3\ii}\phi}{q^{-\ii}\phi^{2\ii}}\right)= \frac{q^{-\ii}\phi^{2\ii}}{q^\ii} A,\nonumber\\
 \Lambda^{-\ii} &=&\frac{q^{\ii}\phi^{-2\ii}}{q^{-\ii}}\left(1+\frac{q^{-3\ii}\phi}{q^{\ii}\phi^{-2\ii}}\right)= \frac{q^{\ii}\phi^{-2\ii}}{q^{-\ii}} \bar{A}.
\end{eqnarray}
We define $f_k=FT(\partial_\theta\ln(f))$ which, taken
together with the definition of the Fourier transform, implies that $f(\theta
+\ii h) = f^{\ii h} \to \e^{-kh}f_k$ provided that the shifted contour lies in the
same analyticity strip, i.e. $\Im(\theta+\ii h) \in (0,2p_0)$. This is
our fundamental domain for $q$ and $\phi$. The $q$ and $\phi$ functions are
related back to the auxiliary functions by taking the difference of the two
Fourier transformed eigenvalues
%\begin{widetext}
\begin{eqnarray}
0 &=& -q_k \e^{-p_0
  k}\left(\e^{k}+\e^{-k}\right)\left(\e^{(p_0-1)k}-\e^{(1-p_0)k}\right) \nonumber \\
&+&
\phi_k\e^{-p_0 k} \left(\e^{(p_0-1)k}-\e^{(1-p_0)k}\right) + A \e^{k} -
\bar{A}_k \e^{-k},\nonumber\\
\label{identityRel}
q_k&=& \frac{\phi_k }{e^{k} +\e^{-k}}+ \frac{\e^{p_0k}\left(A_k\e^{k}-\bar{A}_k \e^{-k}\right)}{\left(\e^{k} + \e^{-k}\right)\left(\e^{(p_0-1)k}-\e^{(1-p_0)k}\right)}. 
\end{eqnarray}
%\end{widetext}
By adding the terms together instead, the largest eigenvalue can likewise be determined in terms of the auxiliary functions as
\begin{eqnarray}
\Lambda_k &=& \phi_k \e^{-p_0 k} \frac{\e^{(2-p_0)k}+\e^{(p_0-2)k}}{e^{k}+e^{-k}}+ \frac{A_k +\bar{A}_k}{\e^k +\e^{-k}}\, .
\end{eqnarray}
The largest eigenvalue is related to the free energy, whereas~\eqref{identityRel} is another form of Eq.~\eq{integralRelation} that is used to determine the Fermi weight for the simple roots of unity. By carrying out the inverse Fourier transform and integrating Eq.~\eqref{identityRel} the relation is explicitly obtained
\begin{equation}
\ln \frac{\phi^{\ii p_0}}{q^{\ii(p_0+1)}q^{\ii(p_0-1)}} = \left[\omega^{\ii} \star \lnb-\omega^{-\ii}\star \lna \right].
\end{equation}
These relations also allow to expand $\frac{\Lambda}{q^{2\ii}q^{-2\ii}}$ into 
%\begin{widetext}
\begin{eqnarray}
  \Lambda_k - q_k\left(\e^{-2k}+\e^{(2-2p_0)k}\right)
%  &=& \phi_k \e^{-p_0 k} \frac{\e^{(2-p_0)k}+\e^{(p_0-2)k}}{e^k+e^{-k}}+ \frac{A_k +\bar{A}_k}{\e^k +\e^{-k}} - q_k\left(\e^{-2k}+\e^{(2-2p_0)k}\right)\nonumber\\
%&=&\frac{A_k \left[\left(\e^{(p_0-1)k}-\e^{(1-p_0)k}\right) -\e^k \left(\e^{(p_0-2)k}+\e^{(2-p_0)k}\right)\right]+ \bar{A}_k \left[\left(\e^{(p_0-1)k}-\e^{(1-p_0)k}\right) +\e^{-k} \left(\e^{(p_0-2)k}+\e^{(2-p_0)k}\right)\right]}{\left(\e^k+\e^{-k}\right) \left(\e^{(p_0-1)k}-\e^{(1-p_0)k}\right)},\nonumber\\
%&=&\frac{A_k \left[-\e^{(1-p_0)k} -\e^{(3-p_0)k}\right]+ \bar{A}_k \left[\e^{(p_0-1)k}+ \e^{(p_0-3)k}\right]}{\left(\e^k+\e^{-k}\right) \left(\e^{(p_0-1)k}-\e^{(1-p_0)k}\right)}=\frac{-A_k\e^{(2-p_0)k} \left[\e^{-k} +\e^{k}\right]+ \bar{A}_k \e^{(p_0-2)k}\left[\e^{k}+ \e^{-k}\right]}{\left(\e^k+\e^{-k}\right) \left(\e^{(p_0-1)k}-\e^{(1-p_0)k}\right)},\nonumber\\
&=&\frac{\e^{(p_0-2)k}\bar{A}_k  - \e^{(2-p_0)k}A_k }{\e^{(p_0-1)k}-\e^{(1-p_0)k}}.\nonumber\\
\end{eqnarray}
%\end{widetext}
This results in the identity used for the eigenvalue trick from the main body of the paper 
\begin{equation}
\ln \frac{\Lambda}{q^{2\ii}q^{-2\ii}} =  \left[\omega^{\ii(2-p_0) }\star\lnb - \omega^{\ii(p_0 -2)}\star \lna\right].
\end{equation}

\section{Proof of Eq.~\eq{fermiWeightEval}}
\label{fermiIdentity}
This proof uses the notation of~Ref.~\cite{kuniba_continued_1998} for
the Takahashi Suzuki (TS) integers and the shorthand for imaginary
shifts $f(\theta+ic) = f^{\ii c}$. Eq.~\eq{fermiWeightEval} comes from
the identification of the particle/hole density of the second to last
string with $K(\theta)$ appearing in the $Y$-system. This
identification appears in the fusion hierarchy of the the easy-plane
($|\Delta|\leq 1$) regime of the XXZ for rational anisotropies
$\gamma$, where $\Delta = \cos(\gamma)$. This section will deal with
the $L-1$-th string again, $\eta(\theta)\equiv\eta_{L-1}(\theta)$,
which is related back to the transfer matrices by
\begin{align}
\eta = K(\theta) = \frac{T_{\mu-1}^{\ii(z_\alpha-z_{\alpha-1}-1)p_0}}{T_{\bar{\mu}-1}^{\ii(y_\alpha + (z_\alpha-z_{\alpha-1}-1)p_0)}}.
\end{align}
The transfer matrices $T_{r-1}(\theta)$ are defined as
\begin{align}
T_{r-1} = q^{\ii r}q^{-\ii r} \sum_{j=1}^r \frac{\phi^{\ii(2j-r-1)}}{q^{\ii(2j-r)}q^{\ii(2j-r-2)}}\, .
\end{align}
Here the notation deviates slightly from~\cite{kuniba_continued_1998} with 
$\phi(\theta) = \left(\sinh\left[\frac{\gamma(\theta+\ii u+\ii)}{2} \right]\sinh\left[\frac{\gamma(\theta-\ii u-\ii)}{2} \right]\right)^{N/2}$, where $N$ is the Trotter number, and the variable $u= -\frac{ J \sin{\gamma}}{T \gamma N}$. The shifts that occur in the transfer matrices are fixed by the anisotropy through the relation
\begin{align}
z_{\alpha-1}p_0 = y_{\alpha-1} + (-1)^{\alpha}p_\alpha = \bar{\mu} + \frac{(-1)^\alpha}{n}.
\end{align}
The equality follows from $y_{\alpha-1} = \bar{\mu}$ and $p_\alpha=1/n$ for an
anisotropy $\pi/\gamma = m /n = [\nu_1, \dots, \nu_\alpha]$ \footnote{Square
  brackets are used to denote continued fractions so that $[2,3] = (2+1/3) =
  7/3$.}, the latter is also used and proven in~\cite{urichuk_spin_2019}. A
second straightforward, but useful equality is that $z_\alpha = n$, so $p_0
z_\alpha = y_\alpha = m$. With these identifications made, the quantity
$\eta +1$ is noted to be the inverse of the Fermi weight for what we
refer to as the spin-particle string, with the $L$-th string being the
spin-hole string. The Fermi weight of the particle string is given by
\begin{align}
\eta+1 = \left(\frac{T_{\mu-1}^{\ii(-\bar{\mu} + (-1)^{\alpha+1}/n -p_0 + m)}}{T_{\bar{\mu}-1}^{\ii(m-\bar{\mu} + (-1)^{\alpha+1}/n -p_0 + m)}}+1\right).
\end{align}
For rational values of $\gamma/\pi$ we have
\begin{align}
\label{specId}
T_{y_\alpha + y_{\alpha-1}-1} = T_{y_\alpha-y_{\alpha-1}-1}+2 T_{y_{\alpha-1}-1}^{\ii y_\alpha}.
\end{align}
For brevity of notation, we set $\dr{w} :=-\bar{\mu} + (-1)^{\alpha+1}/n -p_0 + m$, which can be written as 
\begin{align}
\dr{w} = s-m+\bar\mu+1\, .
% (-\bar{\mu} + (-1)^{\alpha+1}/n -p_0 + m)
\end{align}
with $s$ given in Eq.~\eqref{eq:TSshift}.

Applying Eq.~\eqref{specId}, the inverse Fermi weight may then be
written as
\begin{align}
\eta+1 = \left(\frac{T^{\ii\dr{w}}_{m+\bar{\mu}-1} - T^{\ii(\dr{w}+m)}_{\bar{\mu}-1}}{T^{\ii(m+\dr{w})}_{\bar{\mu}-1}}\right).\label{B7}
\end{align}

Further simplification requires some additional information on the sum
appearing in the definition of the transfer matrix. It is known that
$T_{r-1}^{2m\ii} =T_{r-1}$, which follows from the cyclicity of the $\phi$, $q$
functions. This cyclic relation permits us to conclude that the following sum
is a constant
\begin{align}
\sum_{j=1}^m \frac{\phi^{\ii(2j -1-m)}}{q^{\ii(2j-m)}q^{\ii(2j-2-m)}}=m.
\end{align}
The reason is that all poles of the terms in the sum 
cancel pairwise, they have finite asymptotics, hence the sum is bounded. Due
to Liouville's theorem the sum is constant.
The constant value is identical to the limiting value $m$.

Consequently, the transfer matrices in the numerator of (\ref{B7}) are expanded as
%\begin{widetext}
\begin{eqnarray}
T_{m+\bar{\mu}-1}^{\ii\dr{w}}&=& q^{\ii(m+\bar{\mu}+\dr{w})}q^{-\ii(m+\bar{\mu}-\dr{w})} \nonumber \\
&\!\!\!\!\!\times&\!\!\!\!\! \left( \sum_{j=m+1}^{\bar{\mu}+m} \frac{\phi^{\ii(2j -1-m-\bar{\mu}+ \dr{w})}}{{q^{\ii(2j-m-\bar{\mu}+ \dr{w})}q^{\ii(2j-2-m-\bar{\mu}+ \dr{w})}}} + m\right) \nonumber,\\
T_{\bar{\mu}-1}^{\ii(\dr{w}+m)} &=& q^{\ii(\bar{\mu}+\dr{w}+m)}q^{-\ii(\bar{\mu}-\dr{w}-m)}\nonumber \\
&\times& \sum_{j=1}^{\bar{\mu}} \frac{\phi^{\ii(2j -1+m-\bar{\mu}+ \dr{w})}}{{q^{\ii(2j+m-\bar{\mu}+ \dr{w})}q^{\ii(2j-2+m-\bar{\mu}+ \dr{w})}}}.
\end{eqnarray}
%\end{widetext}
The difference is evaluated by noting that shifts by $2m$ leave the result
unchanged. In the first line $j \to j+m$ is taken, which results in the same
sum as in the second line. Thus the numerator of the inverse Fermi weight
(\ref{B7}) is
\begin{align}
T_{m+\bar{\mu}-1}^{\ii\dr{w}}-T_{\bar{\mu}-1}^{\ii(\dr{w}+m)} = q^{\ii(m+\bar{\mu}+\dr{w})}q^{-\ii(m+\bar{\mu}-\dr{w})} m.
\end{align} 
By expanding all transfer matrices we obtain
\begin{align}
\eta+1 = {m}\left[{\sum_{j=1}^{\bar{\mu}} \frac{\phi^{\ii(2j -1+m-\bar{\mu}+ \dr{w})}}{{q^{\ii(2j+m-\bar{\mu}+ \dr{w})}q^{\ii(2j-2+m-\bar{\mu}+ \dr{w})}}}}\right]^{-1},
\end{align}
or in terms of the particle Fermi weight $\vartheta(\theta)=\frac{1}{\eta(\theta)+1}$
\begin{align}
\vartheta = \frac{1}{m} \sum_{j=1}^{\bar{\mu}} \frac{\phi^{\ii(2j -1+m-\bar{\mu}+ \dr{w})}}{{q^{\ii(2j+m-\bar{\mu}+ \dr{w})}q^{\ii(2j-2+m-\bar{\mu}+ \dr{w})}}}.
\end{align}
This is exactly the relation~\eqref{eq:fermiWeightEval} in the main text.

\section{Amplitudes of next-leading corrections}
\label{FT_App}
For completeness, we reproduce here the amplitudes $Y_2$, $Y_3$, $Y_4$
of the $T^2$ and $T^{4K-4}$ temperature corrections in Eq.~\eqref{FT2}
from Ref.~\cite{pereira_dynamical_2007}.

The bosonized Hamiltonian of the XXZ chain is given by
\begin{eqnarray}
\label{FTH}
H&=& H_0 + H_u + H_{bc}, \quad H_0= \frac{v_0}{2}\int dx\, [\Pi^2+(\partial_x\varphi)^2] \nonumber \\
H_u &=& \lambda \int dx\,\cos(\sqrt{8\pi K}\varphi)] \\
H_{bc}&=&-2\pi v_0\lambda_+\int dx\, (\partial_x\varphi_R)^2(\partial_x\varphi_L)^2 \nonumber \\
&& -2\pi v_0\lambda_-\int dx\, [(\partial_x\varphi_R)^4 + (\partial_x\varphi_L)^4] \nonumber
\end{eqnarray}
where $H_0$ is the standard Luttinger liquid Hamiltonian, $H_u$ the
Umklapp term, and $H_{bc}$ the band curvature terms. The amplitudes
have been determined exactly by a comparison with Bethe ansatz
results \cite{lukyanov_correlation_1999}
\begin{eqnarray}
\label{lambdas}
\lambda &=&
\frac{K\Gamma(K)\sin(\pi/K)}{\pi\Gamma(2-K)}\l[\frac{\Gamma\l(1+\frac{1}{2K-2}\r)}{2\sqrt{\pi}\Gamma\l(1+\frac{K}{2K-2}\r)}\r]^{2K-2},
  \nn \\
\lambda_+ &=& \frac{1}{2\pi}\tan\frac{\pi K}{2K-2} ,\\
\lambda_- &=& \frac{1}{12\pi K}
\frac{\Gamma\l(\frac{3K}{2K-2}\r)\Gamma^3\l(\frac{1}{2K-2}\r)}{\Gamma\l(\frac{3}{2K-2}\r)\Gamma^3\l(\frac{K}{2K-2}\r)} .\nn
\end{eqnarray}
The calculation of the self-energy in second order perturbation theory
in $H_u$ and first order perturbation theory in $H_{bc}$ then yields
the following amplitudes for the next-leading temperature corrections
\begin{eqnarray}
\label{FT_App1}
Y_1&=& \Lambda \frac{B(K,1-2K)}{\sqrt{\pi}2^{2K+1}}\cot(\pi K), \nn \\
Y_2 &=& \Lambda \frac{B(K,1-2K)}{\pi^{5/2}2^{2K+4}}(\pi^2-2\Psi'(K)), \nn \\
Y_3 &=& \Lambda \frac{1}{\pi 2^{4K+4}}\cot^2(\pi K) \Gamma(1/2-K)\Gamma(K) ,\\
Y_4 &=& \frac{\pi^2}{6v^2}(\lambda_+ + 6\lambda_-),\nn\\
\Lambda &=& 4\pi K \lambda^2\sin(2\pi
K)\l(\frac{2\pi}{v}\r)^{4K-2}\Gamma(1/2-K)\Gamma(K), \nn
\end{eqnarray}
with $B(x,y)$ being the Beta function and $\Psi(x)$ being the Digamma
function.

%\bibliographystyle{abbrv} % Tell bibtex which bibliography style to use
%\bibliography{20200301_bibliography,literatur} % Tell bibtex which .bib file to use (this on
%\bibliography{lit_bib} 

\end{document}